\documentclass[%
 aip,
 amsmath,amssymb,
 reprint,twocolumn,%
]{revtex4-1}
\usepackage[english]{babel}

\usepackage{graphicx}
\usepackage{dcolumn}
\usepackage{bm}

\usepackage[utf8]{inputenc}
\usepackage[T1]{fontenc}
\usepackage{mathptmx}
\usepackage{amssymb}
\usepackage{color}
\begin{document}

\preprint{AIP/123-QED}
\title[]{}
\title[]{Roaming leads to amino acid photodamage: A deep learning study of
tyrosine}

\author{Julia Westermayr}
\affiliation{%
University of Warwick, Department of Chemistry, Gibbet Hill Rd, Coventry, CV4 7AL, UK }
\author{Michael Gastegger}
\affiliation{%
Technical University of Berlin, Machine Learning Group, 10587 Berlin, Germany}
\author{Dora V\"{o}r\"{o}s}
\affiliation{%
University of Vienna, Faculty of Chemistry, Institute of Theoretical Chemistry, W\"ahringer Str. 17, 1090 Vienna, Austria.}
\author{Lisa Panzenboeck}
\affiliation{%
University of Vienna, Faculty of Chemistry, Institute of Theoretical Chemistry, W\"ahringer Str. 17, 1090 Vienna, Austria.}
\affiliation{%
Present address: University of Vienna, Faculty of Chemistry, Department of Analytical Chemistry, W\"ahringer Str. 38, 1090 Vienna, Austria.}
\author{Florian Joerg}
\affiliation{%
University of Vienna, Faculty of Chemistry, Institute of Theoretical Chemistry, W\"ahringer Str. 17, 1090 Vienna, Austria.}
\affiliation{%
Present address: University of Vienna, Faculty of Chemistry, Institute of Computational Biological Chemistry, W\"ahringer Str. 17, 1090 Vienna, Austria.}
\author{Leticia Gonz\'{a}lez}
\affiliation{%
University of Vienna, Faculty of Chemistry, Institute of Theoretical Chemistry, W\"ahringer Str. 17, 1090 Vienna, Austria.}
\affiliation{
University of Vienna, Vienna Research Platform on Accelerating Photoreaction Discovery, W\"ahringer Str. 17, 1090 Vienna, Austria.
}
\author{Philipp Marquetand}
 \email{philipp.marquetand@univie.ac.at}
 \affiliation{%
University of Vienna, Faculty of Chemistry, Institute of Theoretical Chemistry, W\"ahringer Str. 17, 1090 Vienna, Austria.}
\affiliation{
University of Vienna, Vienna Research Platform on Accelerating Photoreaction Discovery, W\"ahringer Str. 17, 1090 Vienna, Austria.
}
\affiliation{University of Vienna, Faculty of Chemistry, Data Science @ Uni Vienna, W\"ahringer Str. 29, 1090 Vienna, Austria.
}%

\date{\today}

\begin{abstract}
Although the amino acid tyrosine is among the main building blocks of life, its photochemistry is not fully understood. Traditional theoretical simulations 
 are neither accurate enough, nor computationally efficient to provide the missing puzzle pieces to the experimentally observed signatures obtained via time-resolved pump-probe spectroscopy or mass spectroscopy.
In this work, we go beyond the realms of possibility with conventional quantum chemical methods and develop as well as apply a new technique to shed light on the photochemistry of tyrosine. 
 By doing so, we discover roaming atoms in tyrosine, which is the first time such a reaction is discovered in biology. Our findings suggest that roaming atoms are radicals that could play a fundamental role in the photochemistry of peptides and proteins, offering a new perspective. 
 Our novel method is based on deep learning, leverages the physics underlying the data, and combines different levels of theory. This combination of methods to obtain an accurate picture of excited molecules could shape how we study photochemical systems in the future and how we can overcome the current limitations that we face when applying quantum chemical methods. 
\end{abstract}

\maketitle

\section{Summary}
Light is ubiquitous in our every-day lives and can result in photodamage of molecules with consequences like blindness or skin cancer, but (DNA)-repair processes preventing such harmful reactions also exist. Deciphering the mechanisms that drive these reactions is an extremely complex undertaking and requires a huge effort from experiment and theory.

An especially important molecule, whose photochemistry is not fully understood, is the essential amino acid tyrosine, one of the main building blocks of life. Experimental evidence for photodissociation exists, but a lot of open questions remain. 
Unfortunately, no theoretical method meets all requirements for accurately describing the photodynamics of tyrosine -- and even if a method existed, immense computational costs in the range of several dozen of years of computation would be needed. 

Here, we show how physically-inspired deep learning models can be used to combine the best of different levels of theory to provide an accurate description of the photochemistry of tyrosine.
By doing so, we are able to perform excited-state dynamics simulations beyond the realms of possibility with traditional quantum chemical methods and explore an unexpected reaction pathway, known as roaming. Roaming deviates from common deformation pathways, leads to different photoproducts, and is far from standard chemical intuition. Although unidentified before, we show that it can explain experimentally observed signatures. 

With the discovery of roaming in tyrosine, we reveal a potentially harmful mechanism in biologically relevant systems that could lead to photodamage. The results bring us one step closer toward a better understanding of the fundamental reactions that occur in nature and in our everyday lives, but also raise exciting new questions, \textit{e.g.}, about the importance of roaming in other biological building blocks.

On the one hand, the findings of this work can open new research avenues in the studies of biomolecular systems, where roaming allows for new interpretations and for asking new questions. On the other hand, the deep-learning approach developed in this work shows the possibility to combine different methods and learn from different sources, to study molecules at otherwise unattainable accuracy. 


\section{\label{sec:Introduction}Introduction}
Amino acids form functional peptides and proteins that enable human life on earth and fundamental reactions in nature such as photosynthesis.~\cite{Collini2010N,Cerullo2002s} These systems have been carefully selected to prevent harmful reactions caused by external stimuli, such as UV/visible light. The uptake of UV light leaves the molecule in a highly electronic excited state -- potentially driving toward harmful reactions such as photodegradation, aggregation or bond cleavage.~\cite{Domcke2013NC,Ashfold2006S,Schreier2007S,Rauer2016JACS}  To prevent photodamage, molecules undergo ultrafast, nonradiative transitions from electronic excited states to the ground state on a time scale much faster than irreversible, harmful reactions. Yet, the mechanisms underlying photo-induced reactions in amino acids remain elusive. Thus, knowledge about these mechanisms can substantially contribute to a better understanding of the photostability of peptides and proteins and can further help the design of novel drugs in phototherapy~\cite{Wang2019NC} as well as functional systems with special excited-state properties.~\cite{Marder1997N,Sanchez-Lengeling2018S,Chen2012N}  

Mainly three amino acids are prone to photoexcitation by sunlight: phenylalanine, tyrosine, and tryptophan. Their photodynamics can be studied experimentally, for instance via pump-probe ~\cite{Zewail1994} or high-harmonic spectroscopy,~\cite{Worner2010N} but also theoretically via excited-state dynamics.~\cite{Mai2020ACIE} Photodynamics simulations are very powerful to decipher mechanisms underlying photo-excitation and to provide explanations to experimental observables. While tryptophan, tyrosine, and phenylalanine are studied experimentally,~\cite{Tseng2010PCCP,Gareth2014CS,Iqbal2010JPCL,Tseng2007JPCAb} theoretical simulations of their excited states are extremely expensive, 
such that the smaller chromophores of these molecules (benzene, phenol, and indole) are often the focus of theoretical investigations.~\cite{Sobolewski1999CPL,Gareth2014CS,Oliver2015JPCL,Xie2016JACS} However, size-dependent deactivation pathways suggested by experiments 
question the use of chromophores as model systems to study the photochemistry of the respective amino acids.~\cite{Tseng2010PCCP,Iqbal2010JPCL,Tseng2007JPCAb}
In particular, the photochemistry of tyrosine leaves many questions unanswered. 

Photodissociation of the O-H bond located on the phenol ring (abbreviated as PhO-H in the following) is found to be a major deactivation pathway. Two main dissociation channels, which operate on a slow and a fast time scale, have been proposed for tyrosine and its chromophores after photo-excitation using 200\,nm laser pulses.
~\cite{Iqbal2010JPCL,Iqbal2015} However, a significantly lower signal-to-noise ratio was found in tyrosine, in contrast to its chromophores, phenol and tyramine.  
Existing theoretical studies confirmed a repulsive $^1\pi\sigma^\ast$ state that can lead to photodissociation.~\cite{Tomasello2012JPCB} 
More elaborate theoretical simulations are needed to unravel the excited-state dynamics of tyrosine, but remain computationally infeasible.
Studies have been either limited to static calculations or to low  accuracy.~\cite{Sobolewski2009JPCA,Tomasello2012JPCB} Neither experiments nor theoretical simulations could suggest time constants or a comprehensive picture of the processes that take place in photoexcited tyrosine. 

In this work, we present the first computational method that can predict the excited-state dynamics of tyrosine with high accuracy and on experimentally relevant time scales -- that is on the order of picoseconds with respect to recent experiments.~\cite{Iqbal2015} This is achieved by extending our previously proposed~\cite{Westermayr2019CS,Westermayr2020JPCL} photodynamics approach based on deep neural networks (NNs). We combine different levels of theory, exploit underlying physics when training the NN models as well as introduce boundary conditions. In this way, a new reaction pathway, namely roaming,~\cite{Bowman2011PT,Bowman2011ARPC,Herath2011JPCL,Townsend2004S} for highly excited tyrosine has been discovered 
putting photochemical processes in biology into a new perspective.

\section{Roaming discovered in highly excited tyrosine}
\begin{figure}[ht]
    \includegraphics[width=0.45\textwidth]{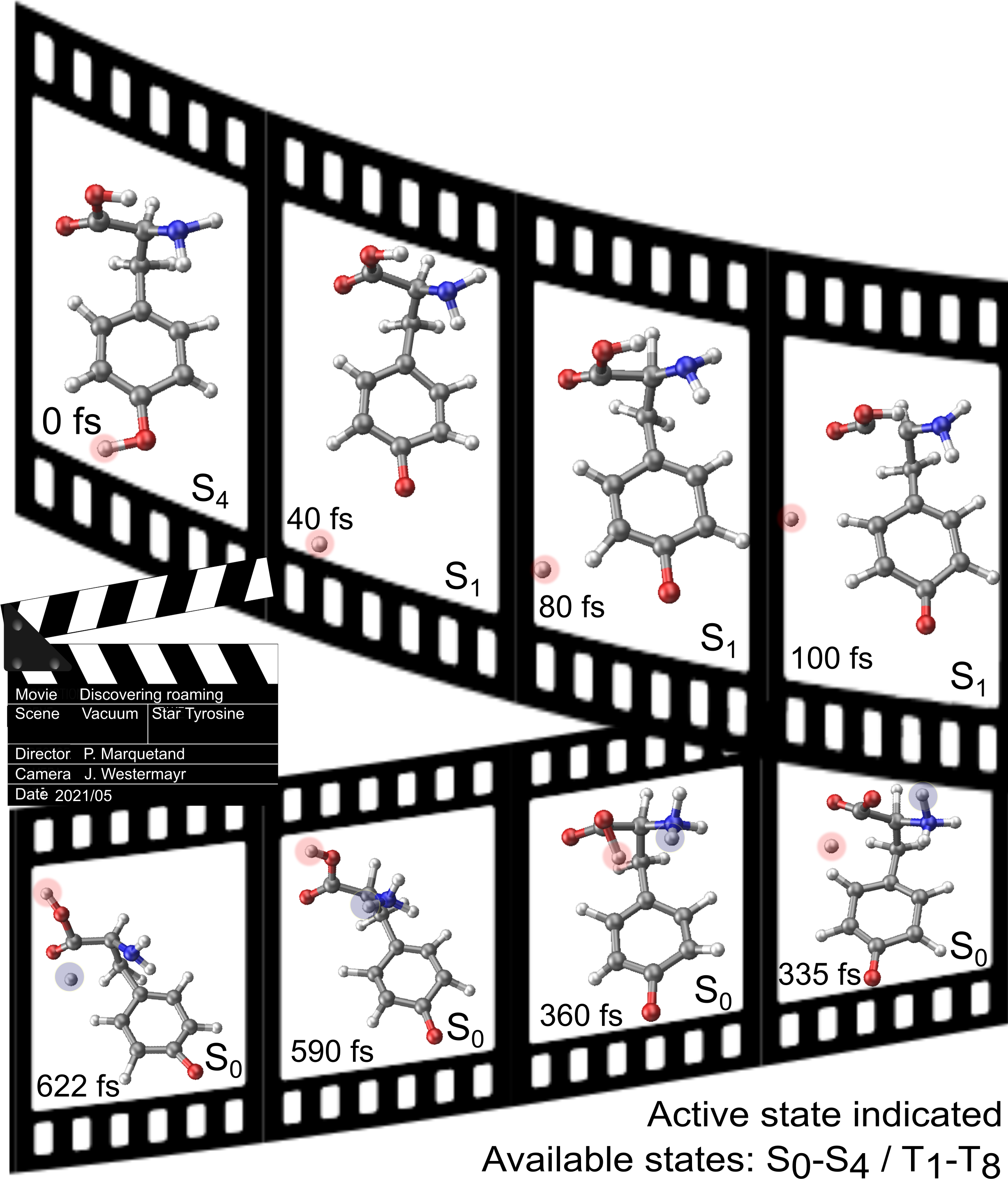}
    \caption{Selected frames of a trajectory showing roaming atoms in tyrosine. The time step and active state are indicated in each frame. }
    \label{fig:roaming}
\end{figure}

Roaming, originally explored in formaldehyde in the pioneering work of Bowman and coworkers from 2004~\cite{Townsend2004S} and actively investigated since then,~\cite{Ekanayake2017SR,Lu2014S,Mereshchenko2015NC,Tso2020SR,Endo2021S,Nandi2021NC}
describes an atom or fragment that moves along a dissociative potential and suddenly changes direction toward an unconventional, dynamically controlled path. The path of roaming fragments deviates from and competes with known deformation pathways and is not what one would expect based on chemical intuition. Characteristics derived from known examples are a large interactomic distance up to 2-3\,{\AA} between the dissociated fragment and the remaining molecule~\cite{Bowman2011PT,Bowman2011ARPC,Herath2011JPCL,Suits2020ARPC} and time scales that can range from hundreds of femtoseconds up to nanoseconds.~\cite{Endo2021S} 
Yet, we are just at the beginning of understanding this special reaction mechanism and its role in nature. It was only in 2020 that roaming fragments have been observed experimentally in real-time.~\cite{Endo2021S}
 
 Experimental studies conducted on tyrosine~\cite{Iqbal2010JPCL,Iqbal2015} 
 are blind to the mechanisms underlying photoexcitation and cannot trace roaming atoms. Another fact that complicates the investigation of the reaction is that, at least in principle, roaming can lead to the same photoproducts. Whether this reaction prevents or promotes photodissociation or leads to long-lived excited states prone to ionization is not known. 
 
Here, we shed light on the dynamical processes that take place after photoexcitation of tyrosine by carrying out dynamics simulations with our SchNarc approach that combines trajectory surface hopping with deep NNs for excited-state properties.~\cite{Westermayr2020JPCL}
The training set for the NNs is to a large part based on the Algebraic Diagrammatic Construction to second order perturbation theory (PT2) method, ADC(2).~\cite{Dreuw2015WIREs} To describe reactions that involve the breaking and formation of bonds, data points based on the complete active space perturbation theory of second order method, CASPT2,~\cite{Roos1980CP,Finley1998CPL} are added.
Corresponding simulations of only 1\,picosecond directly based on the CASPT2 approach would have taken 8 years on a high performance computer but likely would have encountered problematic geometries for the rather small active space (see Fig. S3) and crashed. In this sense, our ML approach offers the unique possibility to carry out such dynamics simulations, while, to the best of our knowledge, neither multi-reference methods nor single reference methods can directly be used in practice to simulate the photodynamics of tyrosine.
A more detailed description on the reference methods can be found in section S1A (Figs S1 and S2 and Tables S1 and S2). The two chosen reference methods are further discussed in section S1B (Fig. S3) and considerations for merging the above described methods are mentioned in section S1C.1. The process of the training set generation and the generation of the artificial data points (Fig. S4 and Tables S3 and S4) can be found in section S1C.2, with the corresponding NN models discussed in section S1G.1 and Figs S10 and S11.
A total number of 29 spin-mixed states, \textit{i.e.}, 5 singlet and 8 triplet states, are learned including the forces as derivatives of the fitted potential energy surfaces and the spin-orbit couplings between the states. Chemical accuracy can be achieved by introducing underlying physics into the NN model. Section S1G.2 provides a detailed explanation of the architecture (Fig. S12) and accuracy (Fig. S13 and Table S6) of the final NN models. 

We simulated over 1,000 trajectories based on the NN potentials to obtain statistical significant results and to discover possible reactions that take place after light excitation.
 We get a first impression of the photodynamics of tyrosine from a representative trajectory in Fig. ~\ref{fig:roaming}, which is also attached as a movie in the supporting information (SI). The active state is indicated at each frame; corresponding excited state potential energy curves are given in Fig. S7. As can be seen, the hydrogen atom that is located at the PhO-H group of the molecule follows a dissociative path and suddenly changes direction at about 80\,femtoseconds to roam around the molecule. In this case, hydrogen abstraction is in competition with the hydrogen transfer from the carboxy group to the amino group of the peptide chain, forming a zwitter-ionic species that is known from recent studies and reflects much better our chemical intuition.~\cite{Tomasello2012JPCB} Interestingly, after hydrogen transfer, the roaming atom attaches to the carboxy group of the peptide chain and the hydrogen atom of the NH$_3^+$ group is transferred toward the phenyl-ring to roam again around the molecule. Roaming is accompanied by internal conversion from the first excited singlet state, S$_1$, to the ground state, S$_0$, which takes place between 300 and 400\,femtoseconds. In contrast, the deactivation from the fourth, bright excited singlet state, S$_4$, to the first excited singlet state is much faster. To make sure that roaming is not an artifact introduced by our NN potentials, we verified this reaction in section S1E using CASPT2 reference calculations (see Fig. S6). 
 
 \subsection{Fragmentation analysis}
  \begin{figure*}[ht]
\centering
\includegraphics[scale=0.5]{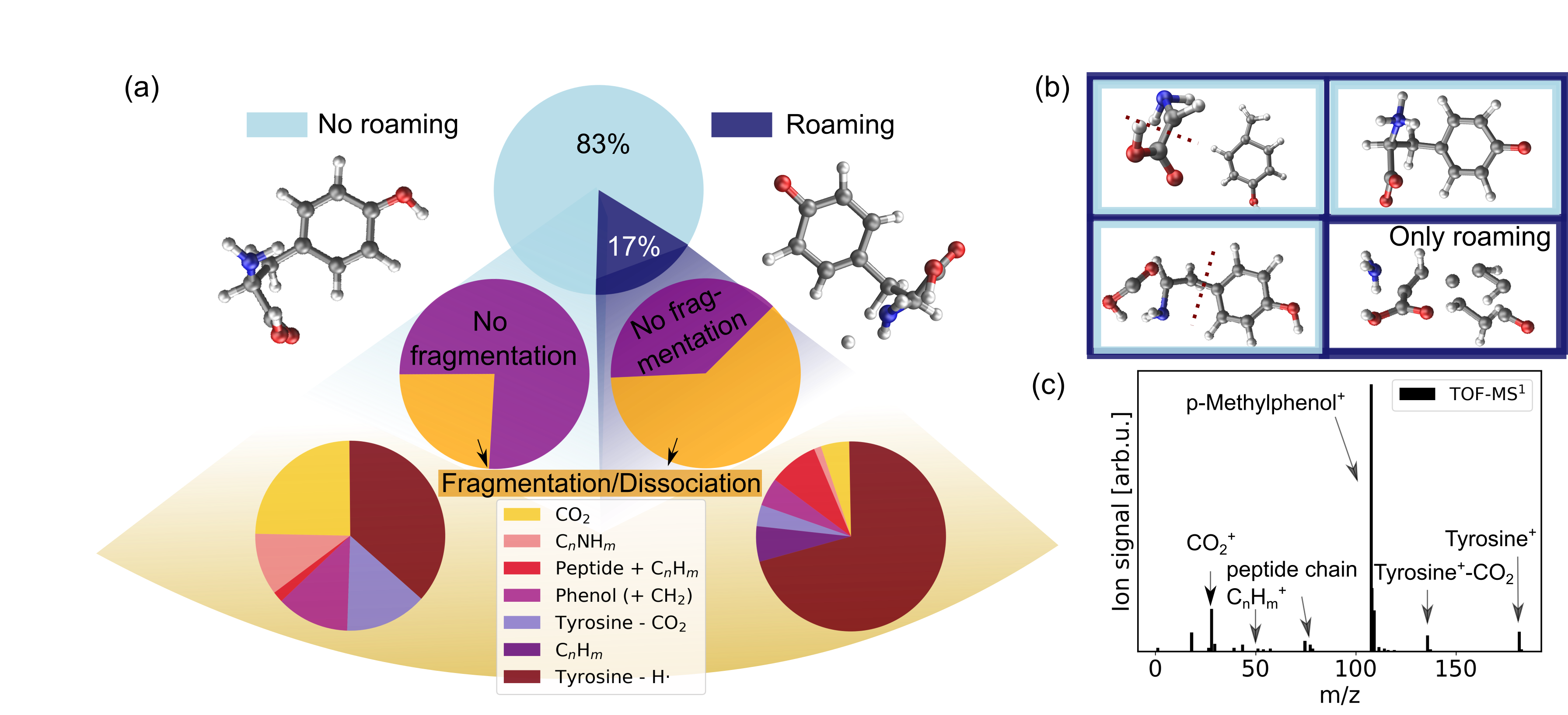}
\caption{(a) Products discovered by the dynamics simulations split into "no roaming" trajectories and "roaming" trajectories. Clustering techniques are used to identify the different types of products in each set of trajectories with (b) representative structures found. (c) Time-of-flight (TOF) mass spectrum extracted from ref.~\citenum{Iqbal2015} that confirms the existence of the discovered reaction outcomes.}
\label{fig:products}
\end{figure*}
 By characterizing of partial charges of 10,767 roaming atoms, we find that roaming atoms are present as radicals. The analysis was carried out with deep NNs trained on dipole moment vectors, which internally form latent partial charges by exploiting the underlying physics of dipole moment vectors.~\cite{Westermayr2020JCP} This workflow needs to be followed as partial charges derived from conventional population analysis, especially those for excited states, would necessitate a tedious recomputation of all conformations sampled during the ML dynamics with quantum chemistry. Details on the model performance and architecture can be found in the SI in sections F2 and G3, respectively, with scatter plots for dipole moments shown in Fig. S14. 

The impact of roaming on the photochemistry of tyrosine was obtained from over 1,000 dynamics simulations that would have been computationally infeasible without the help of deep learning. The simulations were set up according to earlier pump-probe experiments conducted by~\citet{Iqbal2010JPCL} (with further details stated in the SI in section D). Every dynamics trajectory was simulated at least up to 1\,picosecond or until photoproducts were formed. This picosecond time scale was suggested by experimental studies to be sufficiently long to capture all relevant reactions taking place. Nevertheless, a few hundred trajectories were additionally simulated up to 2 and 10\,picoseconds. Fig.~\ref{fig:products} (a) shows the distribution of the products that we split into those obtained from non-roaming and roaming trajectories. As can be seen, we found roaming hydrogen atoms that were originally located at the phenol ring in about 17\% of all trajectories characterized. Due to the large amount of data to analyze, we used k-means clustering (details are specified in section S1F.1 and Fig. S8),~\cite{scikit-learn} to identify different groups of products. 
The findings suggest that molecules that do not show roaming atoms stayed stable during the photodynamics and can prevent most of the harmful reactions that can take place. Only about 10\% of the trajectories show dissociation of a hydrogen atom and 13.5\% show fragmentation within the conducted simulation time. Remarkably, in about 15\% of the cases of hydrogen dissociation, the formation of a zwitter-ionic species can be found (see Fig.~\ref{fig:products} (b) upper right). Regarding the fragmentation process, in non-roaming trajectories, products related to p-methylphenol and peptide chain fragments (see Fig.~\ref{fig:products} (b) upper left) are likely to be generated. Decarboxylation is another possible route that can happen before C-C bond breaking of the peptide chain or after (see lower left example). 
In contrast to non-roaming trajectories, only 36\% of all excited molecules stay stable in roaming trajectories. A large portion of 50\% are characterized by dissociation of the hydrogen atom located at the phenol ring. However, fragmentation cannot be excluded after dissociation and 11\% of dissociated structures result in smaller fragments. In addition, about 14\% of all trajectories show fragmentation without dissociation. The fragments that are found in roaming trajectories are much more diverse. One fourth of all trajectories lead to smaller fragments that are specific to roaming (see lower right picture). Signatures corresponding to these fragments are in agreement to those detected in experiments using multi-mass ion imaging.~\cite{Iqbal2010JPCL,Iqbal2015} The time-of-flight mass spectrum shown in Fig. \ref{fig:products}(c) extracted from ref.~\citenum{Iqbal2015} further confirms the dynamics. Large peaks in the spectrum are related to p-methylphenol, decarboxylation, hydrogen dissociation and hydrocarbons. The findings are in agreement with our current understanding of the impact of roaming on photo-excited molecules.~\cite{Suits2020ARPC} 

\subsection{Kinetics}
\begin{figure}[ht]
    \centering
    \includegraphics[width=0.48\textwidth]{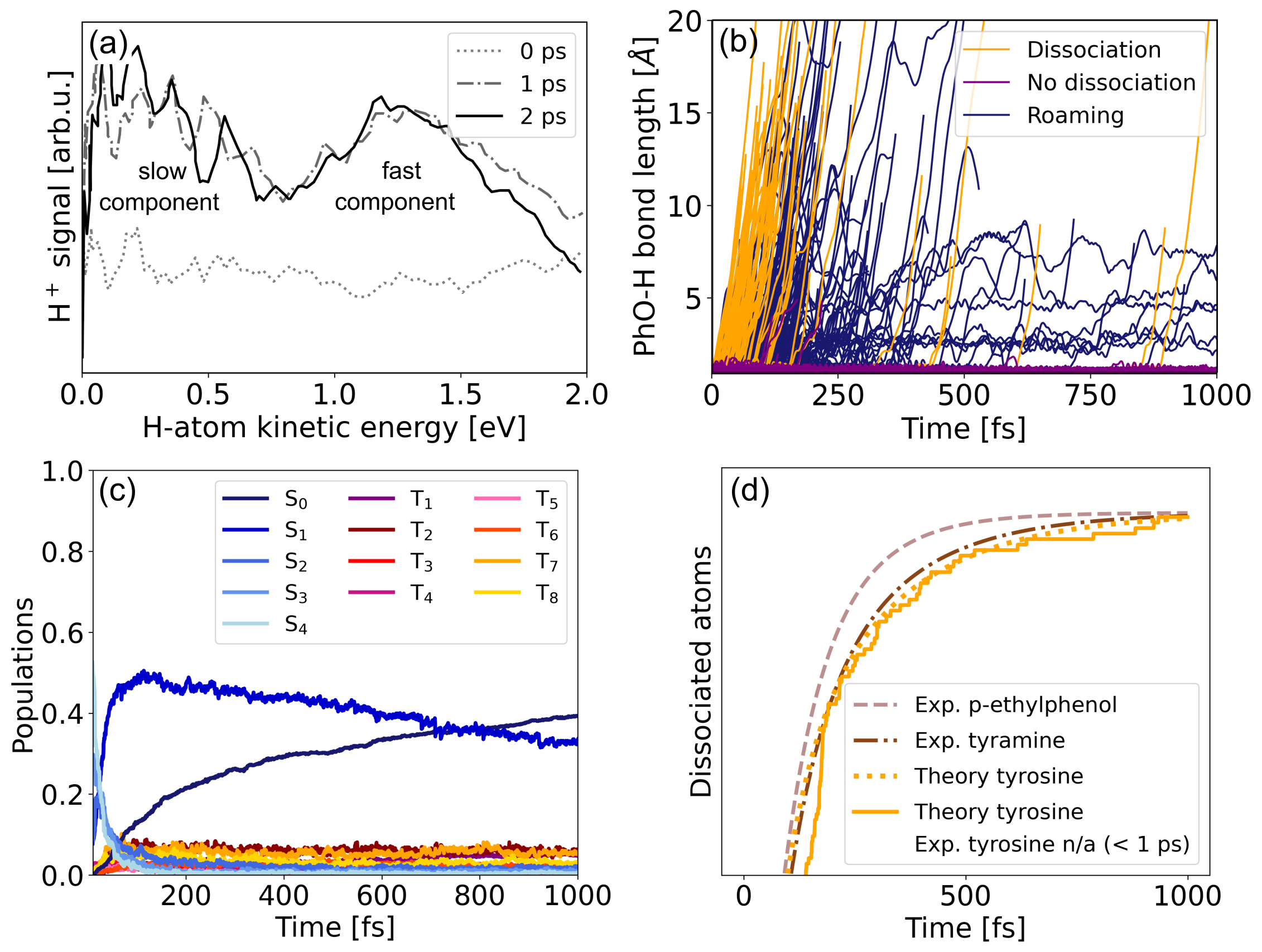}
    \caption{  (a) Kinetic energy release spectrum for tyrosine reproduced from ref.~\citenum{Iqbal2010JPCL}, in which it was obtained via velocity map ion imaging techniques. (b) The bond distance of the hydrogen atom located at the hydroxy group at the phenyl ring (PhO-H) is plotted against simulation time of 1\,picosecond. (c) Population plots averaged from 1022 trajectories simulated up to 1\,picosecond. (d) Amount of trajectories that show dissociation (solid orange) against time. Experimentally found time constants for p-ethylphenol (dashed orange line) and tyramine (dashed-dotted
    purple line) as well as the theoretically fit time constants for tyrosine (dotted orange line) used to fit an exponential function.}
    \label{fig:pop}
\end{figure}

Besides geometrical characterization, the dynamics simulations can be used to obtain population kinetics of the reaction and the impact of roaming thereon. The average populations, after excitation to the S$_4$ state, in the different excited states are plotted in Fig.~\ref{fig:pop}(c). After 1\, picosecond the populations change only slightly, so they are shown up to 10\,picoseconds in Fig. S5 in section S1D.
In contrast, the population transfer to the S$_1$ state is extremely fast and happens on an average of 66\,femtoseconds. After these ultrafast transitions the molecule is found to be stuck in the S$_1$ state. 
About 40\% of the trajectories show population transfer back to the ground state within 1\,picosecond.
 
The pump-probe experiments by Stavros and co-workers~\cite{Iqbal2010JPCL} suggest ultrafast photodissociation and two different reaction channels -- one operating on a slow and one on a fast time scale after excitation with 200\,nm (pump) and probe pulses of 243.1\,nm. The kinetic energy spectrum of tyrosine, extracted from ref.~\citenum{Iqbal2010JPCL}, with the two peaks corresponding to the different kinetic components can be seen in Fig.~\ref{fig:pop}(a). To get an insight into the impact of roaming on the time scales of photodissociation, the PhO-H bond distance of each trajectory is plotted along the simulation time in panel (b). Trajectories are split into those that show roaming and those that do not show roaming -- the latter are additionally split into dissociative and non-dissociative trajectories. The roaming fragments are characterized by bond distances that are mainly within 2-5\,{\AA} to the parent oxygen atom initially. The relatively small distances in the range of 2-3\,{\AA} refer to recoiling hydrogen atoms that bounce against the oxygen atom several times before they either recombine, dissociate or roam around the molecule. An example trajectory of this event is attached as a supporting movie. Noticeably, these large amplitude vibrations were also found in formaldehyde.~\cite{Townsend2004S}

Unfortunately, due to a low signal-to-noise ratio, no experimental time constants are reported for tyrosine. Experiments on smaller chromophores of tyrosine and other amino acids~\cite{Tseng2010PCCP,Iqbal2010JPCL,Tseng2007JPCAb} suggest that the fast component with time constant, k$_1$, results from dissociation in the ground state, whereas the slow component with time constant, k$_2$, is attributed to take place in an excited state. To investigate this assumption, we carry out a two-fold analysis. On one side, the experimentally found time constants for p-ethylphenol and tyramine are used to fit an exponential function, $f(t)$, of the form: $f(t)=a_1\cdot \exp{(-\frac{t}{k_1})}+a_2 \cdot \exp{(-\frac{t}{k_2})}$ with $a_1$ and $a_2$ being constants of 0.15 and 0.1. The function is subtracted from its maximum value and plotted in panel (d). As it is visible, the dissociation is slower in tyramine. In addition, the amount of dissociation found in the dynamics are plotted up to 1\,picosecond (solid dark orange curve), which fits well to the experimental observables. To verify that the slow and fast components are due to dissociation in an excited and ground state, respectively, we split the trajectories into these two categories. The population curves from the S$_1$ to the S$_0$ state are fitted (see Fig. S9 in section S1F.3 in the SI). The fast component has a reaction constant of k$_1$ = 66 $\pm$ 9\,femtoseconds and the slow component has a time constant of k$_2$ = 237 $\pm$ 77\,femtoseconds. These time constants are used to fit the previously defined function $f(t)$ and are shown by dotted lines in panel (d). The results confirm that the two time scales are due to dissociation on different potential energy surfaces.
In addition, the found constants agree very well with the size-dependent kinetics suggested by \citet{Iqbal2010JPCL} and \citet{Tseng2007JPCAb} The reaction rates of p-ethylphenol are reported to be 80 $\pm$ 28\,femtoseconds (k$_1$) and 140 $\pm$ 22\,femtoseconds (k$_2$) and those of tyramine are in the range of 80 $\pm$ 40\,femtoseconds (k$_1$) and 210 $\pm$ 24\,femtoseconds (k$_2$) for the fast and slow components, respectively.  


\section{Conclusion}
In summary, we unravelled the photodynamics of tyrosine by using a combination of different high-level \emph{ab initio} data fitted with deep neural network (NN) potentials. We identified for the first time the distinct characteristics of the photochemistry of this amino acid in agreement with experimental findings.~\cite{Iqbal2010JPCL,Iqbal2015}

Besides the expected photodissociation, we discovered roaming atoms that are far from chemical intuition and compete with other ultrafast deactivation mechanisms. Roaming atoms are characterized by large interatomic distances and deviate from the minimum energy paths. Analysis of latent partial charges obtained from NN models revealed that roaming atoms are present as radicals. While they are found in both, dissociative and non-dissociative trajectories, an analysis with machine learning (ML) clustering models suggest that roaming leads to higher yields of dissociated structures and smaller fragments. Two time components of the dissociation pathway can be distinguished, in line with lifetimes proposed experimentally.~\cite{Iqbal2010JPCL,Iqbal2015} The simulations confirm that the slow and fast time scales originate from dissociation in the ground and excited states, respectively.

The dynamics simulations as well as the analysis could only be achieved with the help of different types of ML methods. The dynamics were conducted with deep NN potentials that were fitted by using combined data of multi-reference methods to capture the bond-breaking and -formation in tyrosine and a single-reference method to provide smooth potential energy surfaces. Due to the complexity of the system and the many states, the underlying physics of the system have been considered, both in the curation of the quantum chemical data as well as in the NN models, to enable the fitting of 29 electronic states and over 1,000 coupling values. Due to the computational efficiency of the prediction of potential energies, derivatives, and couplings provided by the NNs, over thousand trajectories could be simulated on time scales comparable to experiment.

Discovering theoretical evidence for roaming in highly excited tyrosine, one of the main building blocks of life, brings our knowledge one step further toward a better understanding of the photostability and -damage of biological systems. Our results suggest that 
roaming might be a competing relaxation pathway in peptides and proteins, especially in phenomena like hydrogen transfer reactions that are fundamental to nature.~\cite{Marazzi2009JPCL,Shemesh2009JACS}
Yet we are still at the beginning of understanding the secrets behind these mechanisms and their role in nature.

\section*{Data availability}

The molecular coordinates of the used conformers in this study are available as a supplementary file. Additionally, the data set is made available on figshare at 10.6084/m9.figshare.15132081 in the Atomic Simulation Environment (ase)~\cite{Larsen2017IOPP} data base format.
\begin{acknowledgments}
This  work was financially supported by the Austrian Science Fund, W 1232 (MolTag) and the uni:docs program of the University of Vienna (J.W.). The computational results presented have been achieved in part using the Vienna Scientific Cluster. P. M. and L. G. thank the University of Vienna for continuous support, also in the frame of the research platform ViRAPID. J.W. and P. M. are grateful for an NVIDIA Hardware Grant.

\section{Code availability}

\section{Author contributions}
P.M. and L.G. proposed the project and supervised it. J.W. and P.M. implemented and designed the methods. M.G. and D.V. helped fitting ML models. D.V., L.P., and F.J. contributed to reference calculations and the training set generation, \textit{i.e.}, the generation of adjusted data points. J.W. performed the model training, data acquisition and model analysis. J.W. and P.M. interpreted the data, designed the analysis and wrote the initial manuscript. L.G., P.M., J.W., and M.G. revised the manuscript. All authors proofread the final manuscript and supporting information.
\section{Competing interests} There are no competing interests to declare.
\section{Materials \& Correspondence}
\end{acknowledgments}


%
\end{document}


\preprint{AIP/123-QED}
\title[]{}
\title[]{Supporting Information\\ Roaming leads to amino acid photodamage: A deep learning study of
tyrosin}

\author{Julia Westermayr}
\affiliation{%
University of Warwick, Department of Chemistry,Gibbet Hill Rd, Coventry, CV4 7AL, UK}
\author{Michael Gastegger}
\affiliation{%
Technical University of Berlin, Machine Learning Group, 10587 Berlin, Germany}
\author{Dora V\"{o}r\"{o}s}
\affiliation{%
University of Vienna, Faculty of Chemistry, Institute of Theoretical Chemistry, W\"ahringer Str. 17, 1090 Vienna, Austria.}
\author{Lisa Panzenboeck}
\affiliation{%
University of Vienna, Faculty of Chemistry, Institute of Theoretical Chemistry, W\"ahringer Str. 17, 1090 Vienna, Austria.}
\affiliation{Present address: University of Vienna, Faculty of Chemistry, Department of Analytical Chemistry, W\"ahringer Str. 38, 1090 Vienna, Austria.}
\author{Florian Joerg}
\affiliation{%
University of Vienna, Faculty of Chemistry, Institute of Theoretical Chemistry, W\"ahringer Str. 17, 1090 Vienna, Austria.}
\affiliation{%
Present address: University of Vienna, Faculty of Chemistry, Institute of Computational Biological Chemistry, W\"ahringer Str. 17, 1090 Vienna, Austria.}
\author{Leticia Gonz\'{a}lez}
\affiliation{%
University of Vienna, Faculty of Chemistry, Institute of Theoretical Chemistry, W\"ahringer Str. 17, 1090 Vienna, Austria.}
\affiliation{
University of Vienna, Vienna Research Platform on Accelerating Photoreaction Discovery, W\"ahringer Str. 17, 1090 Vienna, Austria.
}
\author{Philipp Marquetand}
 \email{philipp.marquetand@univie.ac.at}
 \affiliation{%
University of Vienna, Faculty of Chemistry, Institute of Theoretical Chemistry, W\"ahringer Str. 17, 1090 Vienna, Austria.}
\affiliation{
University of Vienna, Vienna Research Platform on Accelerating Photoreaction Discovery, W\"ahringer Str. 17, 1090 Vienna, Austria.
}
\affiliation{University of Vienna, Faculty of Chemistry, Data Science @ Uni Vienna, W\"ahringer Str. 29, 1090 Vienna, Austria.
}%

\date{\today}

\maketitle

\tableofcontents

\section{Methods}

The investigation of the photochemistry of tyrosine was enabled with deep neural network (NN) potentials that were fitted by combining two different levels of theory. To the best of our knowledge, machine learning (ML) is one of few, if not the only, viable method to tackle the excited state dynamics of tyrosine on a quantitatively and qualitatively level. This is, because ML can outsource the problem of costly quantum chemistry calculations from the dynamics simulations by pre-fitting the potential energy surfaces from \emph{ab initio} data.~\cite{Behler2021CR,Westermayr2020CR} 
Unfortunately, this is only true if a reference method capable of describing the investigated phenomena is available, which was a challenge for tyrosine. The question that thus had to be addressed and lied at the heart of the fitting is how to combine different theories in the best possible way to enable the photodynamics simulations of tyrosine?

\subsection{Quantum chemistry reference calculations}
Our first step was to choose a proper reference method that could be used to obtain a reliable description of the excited states of tyrosine. To complement experiment, several levels of theory were compared with respect to experimentally available spectroscopic data.

\begin{figure*}[ht]
    \centering
    \includegraphics[scale=0.6]{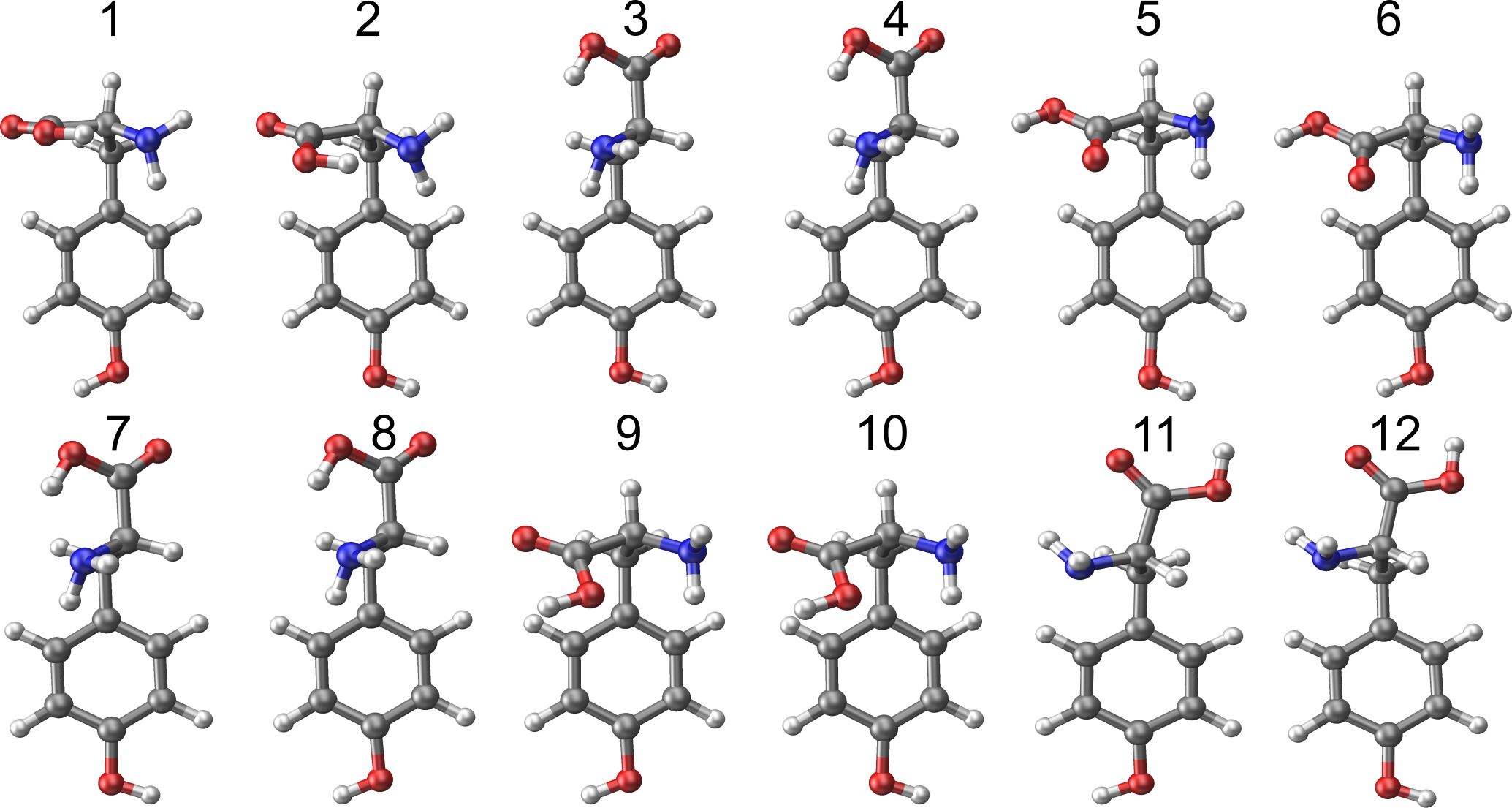}
    \caption{Geometries of the 12 energetically most favorable structures of tyrosine optimized at MP2/TZVP level of theory.}
    \label{fig:conformers}
\end{figure*}

We conducted optimizations of the 12 energetically most favorable conformers of tyrosine. The starting guesses for these structures were extracted from ref.~\citenum{Zhang2005JCP} and optimized using second order M{\o}ller-Plesset perturbation theory (MP2) and the TZVP basis set as implemented in ORCA.~\cite{Neese2012WCMS} The energies of the conformers are reported in Table~\ref{tab:tyrosines} with corresponding structures shown in Fig.~\ref{fig:conformers}. The structures are numbered according to ref.~\citenum{Zhang2005JCP} and can be found in xyz format in a separate supplementary file.~\cite{PHDTHESIS}
\begin{table}[ht]
    \centering
    \begin{tabular}{c|c}
          Conformer Nr. &Energy [kcal/mol]\\
         \hline 
         \hline 
1	&0.00 \\
2	&0.41 \\
5	&0.65 \\
6&0.67	\\
8	&1.27\\
7	&1.29 \\
10	&1.51 \\ 
9	&1.54 \\
3 &2.34 \\
4 &2.35 \\
11	&3.62\\ 
12	&3.72\\ 
    \end{tabular}
    \caption{The energies of the 12 different conformers of tyrosine that were optimized at MP2/TZVP level of theory. Numbering taken from ref.~\citenum{Zhang2005JCP} from which the starting guesses for the different conformers were obtained. Energies are reported with respect to the energetically most favorable structure.~\cite{PHDTHESIS}}
    \label{tab:tyrosines}
\end{table}

The excitation energies of the energetically lowest conformer (conformer 1) are reported in Table \ref{tab:exc} with the different levels of theory that were considered in this work, \textit{i.e.}, the Algebraic Diagrammatic Construction to second order perturbation theory (PT2) method, ADC(2),~\cite{Dreuw2015WIREs} with the cc-pVDZ basis set as implemented in Turbomole,~\cite{TurboMole6.2} the Complete Active Space (CAS) Self-Consistent-Field (SCF) method, CASSCF,~\cite{Roos1980CP} and the multi-state (MS) CASPT2 method~\cite{Finley1998CPL} as implemented in OpenMolcas.~\cite{Galvan2019JCTC} For simplicity, the notation "CASPT2" is used for simulations carried out with MS-CASPT2. CASPT2 and CASSCF rely on an active space, which means that the number of orbitals and electrons that are active during a calculation have to be chosen in advance. The rest of the molecular orbitals, \textit{i.e.}, the energetically lower doubly occupied orbitals as well as the energetically higher doubly unoccupied orbitals, are frozen and not optimized during the calculation. The active space is denoted as CASSCF($n_e$,$n_o$) or CASPT($n_e$,$n_o$) with $n_e$ being the number of active electrons and $n_o$ the number of active orbitals. In addition, experimental values from ref.~\citenum{Smith1929PRSLB} and literature values obtained from Time-Dependent Density Functional Theory (TD-DFT) with the CAM-B3LYP hybrid functional~\cite{Yanai2004CPL} and the SVP basis from ref.~\citenum{Tomasello2012JPCB} are reported. As can be seen, the energies of CASPT2(12,11) and ADC(2) are relatively close to each other, whereas CASSCF(12,11) overestimates the excitation energies of excited singlet and triplet states. All methods overestimate experimentally found excitation energies. 
\begin{table}[h]
    \caption{Excitation energies~\cite{PHDTHESIS} of conformer 1 with respect to Table~\ref{tab:tyrosines} and Fig.~\ref{fig:conformers} of tyrosine. 5 singlet states and 8 triplet states are computed with ADC(2)/cc-pVDZ, CASSCF(12,11)/ano-rcc-pVDZ, and CASPT(2)/ano-rcc-pVDZ. Values from literature are additionally included and refer to TD-DFT computation with CAM-B3LYP/SVP~\cite{Tomasello2012JPCB} and to experiments from ref.~\citenum{Smith1929PRSLB}.}
    \centering
    \begin{tabular}{c|ccccc}
         State& ADC(2) & CASSCF(12,11) &CASPT2(12,11) &TD-DFT~\cite{Tomasello2012JPCB} &Exp.~\cite{Smith1929PRSLB}\\
         \hline
         \hline
       S$_1$  & 5.00	& 4.94	& 4.63 &5.20&$\sim$ 4.43\\
S$_2$& 5.92	& 6.43	& 5.99 &-&$\sim$ 5.39\\
S$_3$& 6.20	& 7.87	& 6.24 &6.07\\
S$_4$& 7.16	& 8.19	& 7.70&6.92\\
T$_1$& 4.27	& 3.86	& 3.91\\
T$_2$& 4.60	& 4.91	& 4.34\\
T$_3$& 5.01	& 4.98	& 4.49\\
T$_4$& 5.58	& 6.10	& 5.32\\
T$_5$& 6.19	& 6.94	& 5.65\\
T$_6$& 6.75	& 7.14	& 7.01\\
T$_7$& 6.99	& 7.31	& 7.10\\
T$_8$& 7.03	& 7.47	& 7.15\\
    \end{tabular}
    \label{tab:exc}
\end{table}{}

\begin{figure}[ht]
    \centering
   \includegraphics[width=0.45\textwidth]{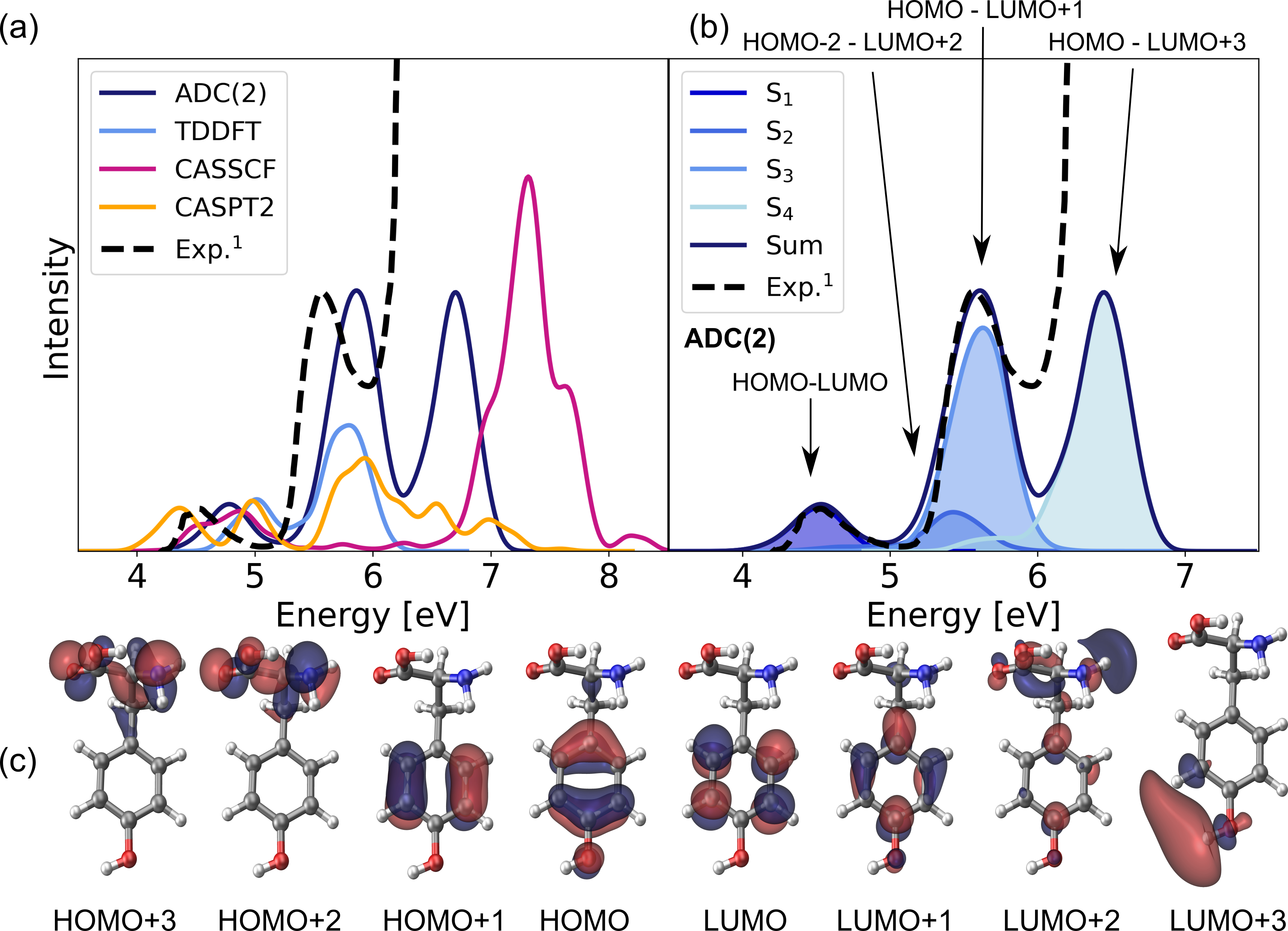}
   \caption{(a) The absorption spectrum of tyrosine computed with ADC(2) for 1000 Wigner-sampled conformations is shown together with the experimental spectrum for tyrosine in the gas phase ("Exp.") obtained from ref.~\citenum{Smith1929PRSLB}. TD-DFT/PBE0/SV(P), CASSCF(12,11), and CASPT(12,11) spectra are additionally shown using 100 Wigner-sampled conformations. The full width at half maximum for the Gaussian convolution was 0.2\,eV in all theoretical spectra. The absorption peaks are scaled according to the energetically lowest lying experimental absorption peak. (b) The spectrum obtained from ADC(2) is shifted in energy by 0.25\,eV and contributions from different excited states are shown. The corresponding vertical excitation energies are specified in Table \ref{tab:exc}. (c) The minimum energy conformer of tyrosine along with molecular orbitals around the HOMO and LUMO computed for 5 singlet states using ADC(2) are shown. Main contributions of orbitals for each singlet state are indicated in panel (b).}
    \label{fig:methodssi}
\end{figure}

The UV/visible absorption spectra are calculated from Wigner-sampled \cite{Wigner1932MM,Wigner1932PR} conformations with ADC(2)/def2-SVP (1000 sampled structures), TDDFT/PBE0/SV(P) (100 sampled structures), CASSCF(12,11)/ano-rcc-pVDZ (100 sampled structures), and CASPT2(12,11)/ano-rcc-pVDZ (100 sampled structures) and are shown in panel Fig. \ref{fig:methodssi}(a).
It can be seen that ADC(2) can reproduce the experimentally measured spectrum of tyrosine, while other methods such as CASPT2(12,11), CASSCF(12,11) or TDDFT as used in literature~\cite{Tomasello2012JPCB} exhibit pronounced differences. 
The contribution of each singlet state obtained from ADC(2) to experimental UV/visible absorption peaks can be seen in Fig.~\ref{fig:methodssi} (b) after shifting of ADC(2) values by 0.25\,eV. From the relevant orbitals in panel (c) and their contributions to the respective states marked in panel (b), it is visible that the PhO-H bond is relevant for the fourth excited singlet state and also orbitals located at the peptide group, \textit{i.e.}, the amino and carboxy group. Based on previous TDDFT~\cite{Tomasello2012JPCB} and experimental~\cite{Iqbal2010JPCL,Iqbal2015} studies as well as scans along different reaction coordinates with ADC(2), CASSCF, and CASPT2, these orbitals should be important for the excited state dynamics. Worth mentioning is that hydrogen transfer from the neutral form of tyrosine to the zwitter-ionic form, structural changes in the peptide group, and hydrogen dissocation reactions were reported. Orbitals located at the amino group could not be described using CASPT2 or CASSCF even with larger active spaces than (12,11). Further, the UV/visible absorption spectra obtained from these methods do not match the experimentally found spectrum of tyrosine. As a result, dynamics simulations are likely to fail with CASSCF or CASPT2. 

In fact, we evaluated several active spaces along with a different number of active states for CASSCF and CASPT2 up to the point until calculations became too expensive, \textit{i.e.}, up to 14 electrons in 18 orbitals.   One calculation of 6 singlet states without forces took about 5 hours and 40 minutes on a 2x Intel Xeon E5-2650 v3 central processing unit (CPU), making larger active spaces practically infeasible for thousands of data points with more electronic states and additional force calculations. Hence, 12 active electrons in 11 active orbitals in combination with the basis set ano-rcc-pVDZ~\cite{Roos2004JPCA} provided a good compromise between efficiency and accuracy. With all tested active spaces and active states, we were not able to cover all relevant orbitals. Based on literature, ADC(2) calculations (see panel (c) in Fig.~\ref{fig:methodssi}), and chemical intuition, these were considered to be the $\pi$/$\pi^\ast$ orbitals of the phenyl and carboxyl group, the n-orbital of the lone pair of the nitrogen atom and $\sigma$/$\sigma^\ast$ orbitals of the hydroxy group at the phenyl ring or at the amino group. We found that the lone pair of the nitrogen atom of the amino group was missing in all conducted active-space calculations. However, this orbital plays a role in the excited states according to ADC(2) and according to TDDFT.~\cite{Tomasello2012JPCB}  Consequently, the resulting CASSCF and CASPT2 potential energy curves along different reaction coordinates were not smooth (see Fig.\ref{fig:methods}(a) below for an example), posing a problem to ML models.

Due to the aforementioned considerations and the method comparison, the ADC(2) method was chosen for generating the largest part of the training set for ML training. As can be seen in the UV/visible absorption spectra in Fig.~\ref{fig:methodssi}(b) and from Table~\ref{tab:exc}, the ADC(2) method is in very good agreement to experiment. We note that ADC(2) is based on restricted MP2 in the calculations reported in this study. Although homolytic bond breaking is tackled, unrestricted MP2 as a reference for ADC(2) brought no advantage since the perturbation terms became too large in the dissociation limit in both MP2 variants. With ADC(2) based on restricted MP2, spin contamination could be avoided and the spin-orbit couplings (SOCs) between the first and second triplet states could be computed reliably, which was not the case with ADC(2) based on unrestricted MP2.


\subsection{Training set generation}
The training set of tyrosine was computed with ADC(2) for 5 singlet states and 8 triplet states and contains energies, forces, and SOCs between singlet and triplet states and triplet and triplet states.
Starting from the most stable conformer of tyrosine, unrelaxed reaction scans were carried out along different normal modes and combinations of normal modes. In addition, a scan from the neutral to the zwitter-ionic form was included in the training set. 698 data points were collected in this way. To obtain a larger initial training set, scans along different normal modes and combinations of normal modes of two different conformers of tyrosine (conformers 5 and 7 in Fig.~\ref{fig:conformers}) were included in the training set. These additional conformers contributed respectively with 622 and 647 data points, resulting in a total number of 1,967 data points. 

In order to fit SOCs with ML, the phase correction procedure previously reported in ref.~\citenum{Westermayr2019CS} was applied, which corrects the arbitrary sign jumps that are due to the arbitrary phase of the wave function. These arbitrary jumps need to be accounted for in the training set by preprocessing data. In addition, a phase-free training algorithm could be used to account for arbitrary sign jumps during the training process. The phase free training algorithm computes the error of a prediction, $Y^{ML}$, to the original value, $Y^{Ref.}$, by computing all possible phase vectors between electronic states and taking the minimum error. 
With 13 different spin-diabatic states, 4,096 different phase-vector combinations exist, making the phase-free training process expensive. Therefore, we resorted to phase correction of the training set, which makes the training set generation costly. This procedure required one reference calculation. In this work, we chose the equilibrium structure for this purpose. A wave function overlap computation from this reference data point to every other data point that was included in the training set was carried out. If the overlap of two states was larger than the absolute value of $\pm 0.5$, the corresponding property between these states could be corrected. If the overlap was smaller, interpolation steps between these geometries were required. Hence, the larger the explored conformational region of tyrosine became, the more expensive the phase correction procedure turned out to be. Further, the higher the density of states was at a given configuration, the more electronic states had to be computed, making some parts of the conformational space untraceable. We solved this issue in a later stage of the training set generation using ML models for finding the correct phase vector. We did so by predicting the affected properties with an ML model trained on phase-corrected data. The phase vector that gave the lowest error from the predicted to the reference property was used to correct the data. In the early stages of training set generation when the model was not yet capable of predicting properties affected by the arbitrary phase, we excluded all data points that could not be corrected with the computation of one singlet and one triplet state more than used in the training set (in total 6 singlets and 9 triplets). 


The initial training set was further extended using an active learning technique, namely adaptive sampling.~\cite{Behler2015IJQC,Gastegger2017CS,Westermayr2019CS} This method made use of ML models that were trained on the initial, incomplete training set. Dynamics simulations with these ML models were initialized. At every time step, the predictions of two to six ML models that differed slightly in their hyperparameters were compared to each other. If the ML models differed strongly from each other, it was assumed that an unexplored or undersampled conformational region of the molecule was reached. In this way, a training set for dynamics simulations could be built comprehensively and computationally efficiently. In more detail, this procedure was carried out with 25 different initial conditions (coordinates + velocities) of conformer 1 and 5, which were generated from the respective Wigner distributions.~\cite{Wigner1932PR} Conformer 5 was chosen as it differed to conformer 1 by a rotation of the carboxy and hydroxy group and was still more similar to conformer 1 than the other conformer, namely conformer 7, which was included in the intitial training set. This should provide an initial training set that could be used to propagate the molecule after light excitation at least for a short time scale and to allow for accurate enough potentials to expand the training set further via adaptive sampling.  Each initial conformation was excited to the S$_1$, S$_2$, S$_3$, and S$_4$ state, respectively, resulting in 100 trajectories per conformer. Photodynamics simulations were carried out with initial multi-layer feed-forward NN potentials using the inverse distance matrix as a descriptor for the geometries. The different NN models used in this study are specified in more details in section~\ref{sec:ml}. 

The molecular dynamics program that we applied for adaptive sampling was SHARC (Surface Hopping including ARbitrary Couplings).~\cite{sharc-md2,Mai2018WCMS} 
At each time step during the dynamics, an average of 6 NN predictions for energies, forces, and SOCs (where the first run employed only the average of 2 NNs) was used for propagating the nuclei. At each time step of 0.5\,femtoseconds, the predictions of the networks were compared to each other. Whenever the energies, forces, and SOCs predictions deviated by more than 25\,kcal/mol, 
25\,kcal/mol/a.u., or 2.2$\cdot$10$^{-5}$\,a.u., respectively, simulations were stopped. These thresholds can be considered as rather large, which we chose on purpose in order to not get stuck in already sampled regions with the initial models that could also suffer from the little data used. 
The molecular geometries at the points at which the trajectories were terminated were recomputed with the reference method and were added to the training set if phase-correction was successful. We deemed a phase correction to be successful if all phases of the electronic states could be traced back to the initial reference geometry with high confidence, \textit{i.e.}, if the overlap between two states of two interpolated geometries exceeded 50\%. All interpolation steps carried out during the phase correction were further added to the training set. The NNs were then re-trained using the expanded training set and the pre-defined thresholds for NN comparisons were multiplied with a factor of 0.95 in each iteration. This procedure was carried out with all 100 trajectories of a conformer simultaneously until the number of data points approximately doubled. The network architecture was re-optimized afterwards and adaptive sampling was restarted from the initial conditions. 

During the adaptive sampling dynamics simulations with NNs, the nonadiabatic couplings, which enter the equations to compute the probabilities for internal conversion in SHARC, were set to 0. Hence, no hops between states of same spin-multiplicities were allowed, which was assumed reasonable in this exploratory run, since trajectories were launched on all considered excited singlet states. Still, intersystem-crossing between states of different spin-multiplicities or between triplet states were allowed during the dynamics simulations. 16,738 data points were obtained as an initial training set. 

The adaptive sampling runs were carried out with multi-layer feed-forward NNs as described in ref.~\citenum{Westermayr2019CS} in combination with the matrix of inverse distances, while the actual dynamics simulations were done with SchNarc models as described in ref.~\citenum{Westermayr2020JPCL}. The reason is that the former NNs are more efficient at the training stage while the latter NNs are more efficient at the prediction stage.
The training of the standard feed-forward NNs could be carried out in approximately 4-16 hours depending on the training set size on a single CPU. 
For comparison, the training of SchNet models for excited states~\cite{Westermayr2020JPCL} took 3 days and 14 days on a Tesla-V100 graphic processing unit (GPU) for energies/forces and SOCs, respectively, using 15,000 data points. 

\subsection{Additional data}\label{sec:artificial_data}
One physical aspect that ADC(2) is not able to predict  is that several pairs of singlet and triplet states should become degenerate in the dissociation limit. Therefore, \textit{ad hoc} data points were introduced that respected these physical relations.  The created \textit{ad hoc} data points were derived from CASPT2 calculations. The decision to combine these two methods is justified and discussed below.

\subsubsection{Considerations for merging single-reference data with manipulated data based on multi-reference methods}
%

\begin{figure}[ht]
    \centering
   \includegraphics[width=0.45\textwidth]{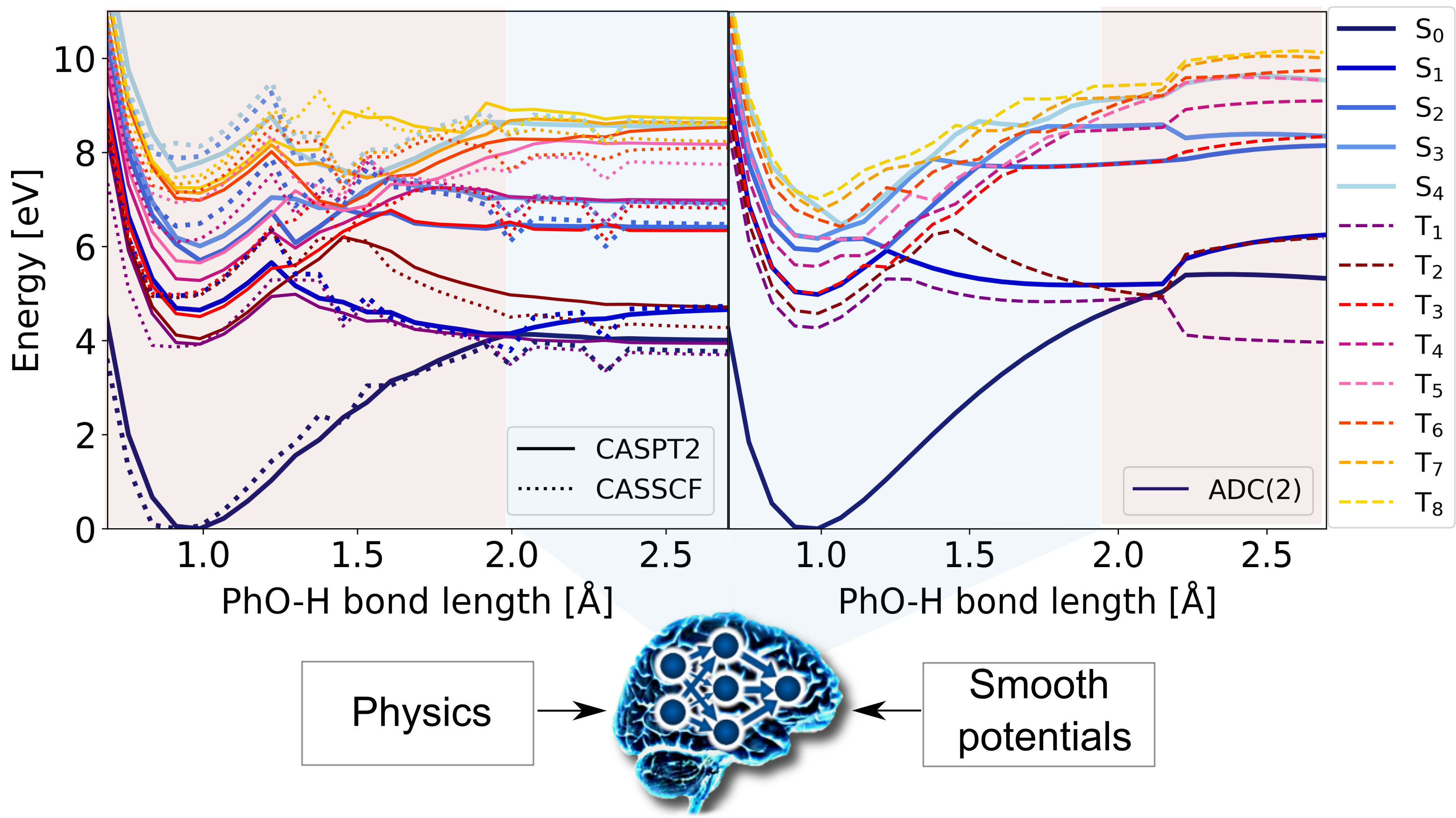}
    \caption{Potential energy curves along the O-H bond located at the phenyl ring (PhO-H) using CASSCF(12,11) (left panel), CASPT2(12,11) (left panel), and ADC(2) (right panel).}
    \label{fig:methods}
\end{figure}

As already mentioned, CASSCF and CASPT2 require an active space that includes all relevant orbitals for reactions taking place during the photodynamics simulations. Besides their computational costs (a single simulation during 1\,picosecond would take up to 8 years on a high performance computer with CASPT2(12,11)), the simulations would likely crash before reaching the desired time scale due to inconsistencies in the active space along different trajectories. As can be seen in the dissociation curves in Fig.~\ref{fig:methods} in the left panel, states were likely to switch their character along a reaction path leading to inconsistent potential energy curves. In addition, trials to fit potential energy surfaces obtained from CASSCF or CASPT2 failed with ML methods. This failure could be attributed to the many excited states of tyrosine that were energetically close to each other. The high density of states complicated the use of multi-reference methods. Further, significant jumps in the potential energy prohibited a meaningful fitting with ML.~\cite{Westermayr2020CR}

In contrast to highly accurate multi-reference methods, ADC(2) provided smooth potential energy curves, which can be seen in Fig.~\ref{fig:methods} in the right panel.
However, ADC(2) came with a decrease in accuracy and in fact, it is less suitable to describe strongly correlated systems such as tyrosine,\cite{Westermayr2020CR,Crespo-Otero2018CR} especially for bond-breaking and -formation.~\cite{Giesbertz2008CPL} Consequently, the potential energy curves at large interatomic distances were qualitatively wrong and showed a splitting of singlet and triplet states, which should have been degenerate at the dissociation limit.

We therefore concluded that neither multi-reference methods, nor single reference methods could be used in practice to simulate the photodynamics of tyrosine.
However, ADC(2) was found to give a reasonably accurate description of tyrosine in non-dissociative regions (see also the absorption spectra obtained with ADC(2) in Fig. \ref{fig:methodssi} compared to the experimental spectrum) and CASPT2 could describe the correct asympthotic behaviour. Thus, we resorted to exploit the expressive power of deep NNs to combine the best of both worlds: 
1) The smooth potential energy surfaces for singlet and triplet states including their derivatives and SOCs were provided from the ADC(2) method.~\cite{Dreuw2015WIREs} 2) The correct description of bond breaking and formation was fitted using CASPT2 reference data. A total number of 29 spin-mixed states, \textit{i.e.}, 5 singlet and 8 triplet states, were learned. Chemical accuracy could be achieved by introducing underlying physics into the NN model. This was achieved by learning SOCs in two different representations and in combination with energies. Details on the implementation are given in the SI in section \ref{sec:ml} including the training process and the validation of the model. 


\subsubsection{Generation of artificial data points}
\begin{figure*}[ht]
    \centering
    \includegraphics[scale=0.55]{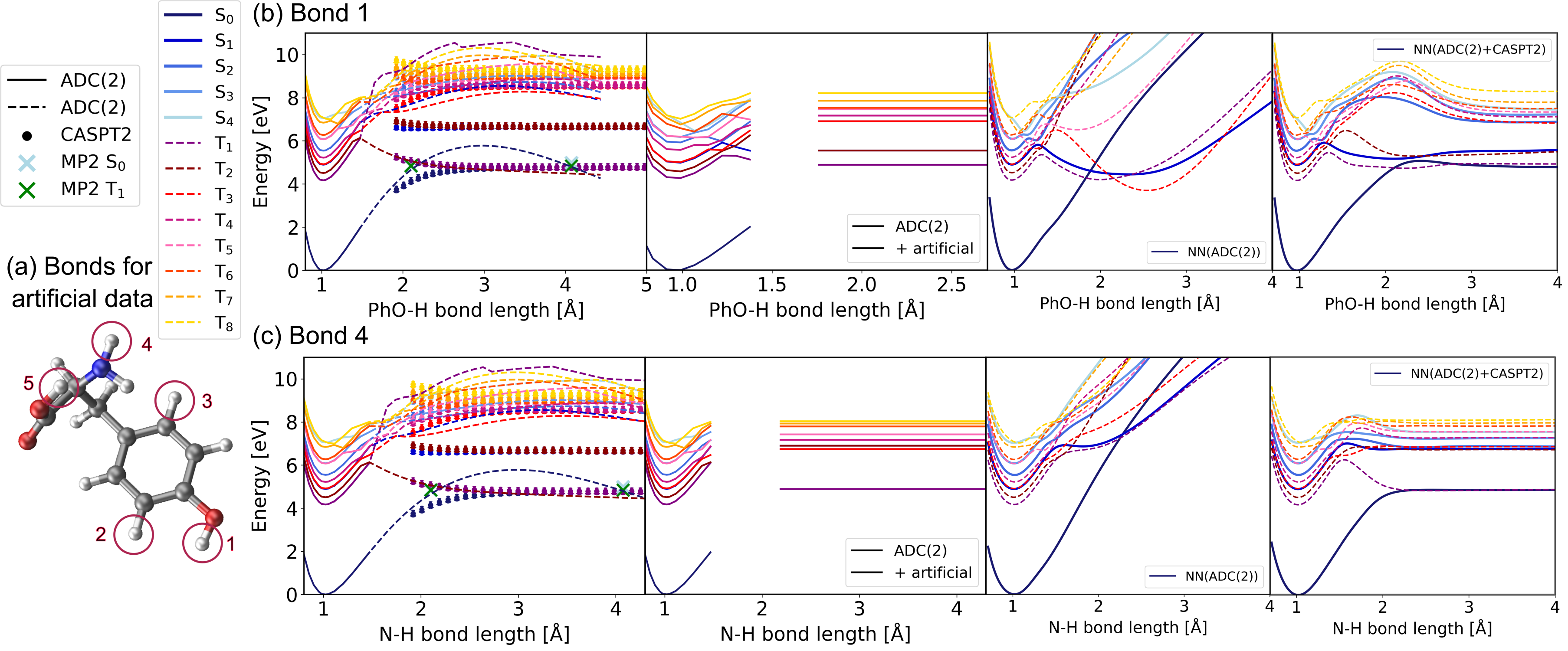}
    \caption{ (a) Hydrogen atoms of tyrosine that are circled were used to generate artificial data by elongation of the respective X-H bond with X=C,N,O. (b) The PhO-H bond length and (c) the N-H bond length as example reaction coordinates to show the differences of ADC(2) (solid and dashed lines) and CASPT2(12,11) (circles) potential energies in the leftmost plots. MP2 calculations (crosses) manifest the degeneracy of singlet and triplet states. The second plot illustrates the data used for training with ADC(2) and artificially generated data shown by solid lines. The last two plots show NN predictions that were trained without and with manipulated data points, respectively. S refers to singlet states and T to triplet states.}
    \label{fig:diss}
\end{figure*}

105 data points were added from 5 scans along different dissociation coordinates using the energetically lowest lying conformer of tyrosine. The corresponding hydrogen atoms are indicated in Fig.~\ref{fig:diss} (a) and were initially attached at the hydroxy group located at the phenyl-ring (1), at two C-atoms located at the phenyl ring (2 and 3), at the amino group (4), and at the carboxyl group (5). The hydrogens were detached along the vectors in which their bonds originally pointed. Unrelaxed scans were carried out, \textit{i.e.}, the rest of the atoms did not change position. The reaction coordinates were chosen with the aim of including a dissociation from every heavy atom type (C, N, O) and based on literature, where for example a hydrogen transfer was reported from the carboxyl group or from the amino group and to the carbon atom of the phenyl-ring.~\cite{Tomasello2012JPCB}
To check the maximum bond length up to which ADC(2) provided valid reference data, we recalculated several points along the dissociation scans using CASPT2(12,11) and MP2. Comparison of the energy values from these different theory levels led us to define different maximum bond lengths up to which the ADC(2) data were usable along the different X-H coordinates. These maximum distances for the use of the ADC(2) data, $r_{max}^{\text{ADC(2)}}$, are given in Table~\ref{tab:diss} along with the respective X-H equilibrium distances.

{}

As an example for the artificial data generation, the potential scans along the PhO-H bond and the N-H bond of tyrosine are given in Fig.~\ref{fig:diss} (b) and (c). In the leftmost panels, ADC(2) scans are illustrated using solid lines up to the molecular geometry that was included in the training set. Noticeably, at that point, also the phase correction algorithm~\cite{Westermayr2019CS} could not be used anymore, because states that were very high in energy at the equilibrium geometry entered the range of considered states here, so that their phases could not be tracked reliably. The dashed liss indicate ADC(2) results that were not included in the training set. CASPT2 energies are shown by dots and unrestricted MP2 energies by blue and red crosses. The absolute energy at the respective equilibrium distance was set to zero for all methods. The ADC(2) training data and the artificially fitted training data are illustrated with solid lines in the second left-most plots of panels (b) and (c).

As can be seen in the leftmost panels, CASPT2(12,11) describes the dissociation event accurately and shows that at large distances, the S$_0$/T$_1$, S$_1$/T$_2$, S$_2$/T$_3$, S$_3$/T$_4$, and S$_4$/T$_5$ states become degenerate. In contrast, ADC(2) incorrectly describes this event and shows a splitting of singlet and triplet energies. To include dissociated geometries in the training set, the relative energies of the \textit{ad hoc} data points were adjusted from CASPT2(12,11) calculations to the MP2 energy of the ground state (since the ground state in ADC(2) was taken from MP2 calculations) and scaled to the energy range obtained in the ADC(2) calculations at r$_{max}^{\text{ADC(2)}}$. The bond distances that were added manually to the training set are given in Table~\ref{tab:art}. In more details the data points were manipulated as follows: In a first attempt, the degeneracy of the respective states (S$_0$/T$_1$, S$_1$/T$_2$, S$_2$/T$_3$, S$_3$/T$_4$, and S$_4$/T$_5$) at large bond distances was confirmed with CASPT2(12,11) calculations for each scan separately. Afterwards, several unrestricted MP2 calculations were carried out. At large bond distances, unrestricted MP2 also shows that the S$_0$ and T$_1$ states are degenerate. The energies at this point were taken for the artificial data points for the S$_0$ and T$_1$ states. The energies of T$_5$, T$_6$, and T$_7$ were left unchanged from ADC(2) calculations at the point $r_{max}^{\text{ADC(2)}}$. The remaining energies were obtained by taking the energy difference from CASPT2(12,11) calculations of each pair of degenerated states, \textit{i.e.} S$_1$/T$_2$, S$_2$/T$_3$, S$_3$/T$_4$, and S$_4$/T$_5$, to the S$_0$/T$_1$ pair.
The SOC values were kept constant from $r_{max}^{\text{ADC(2)}}$. The forces were first taken from the equilibrium geometry since an unrelaxed scan was carried out. Only the forces along the dissociation coordinate are then set to zero and the remaining ones adapted from the original forces by applying Gram-Schmidt orthogonalization. 
 
The third and fourth plots of panels (b) and (c) show the NN potentials that were obtained from the previously generated training set of 16,654 data points without and with artificial data points, respectively. As can be seen, models without manipulated data that reflect physically correct behaviour fail to reproduce dissociation accurately. In contrast, the NN potentials that were additionally trained on dissociated geometries can reproduce this reaction, as expected. This training set was deemed sufficiently large to train SchNet models and conduct photodynamics simulations. To ensure accurate dynamics, we included another adaptive sampling run that will be discussed in the next section. 

\begin{table}[h]
    \centering
        \caption{\label{tab:art}Bond distances between X and H (X=C,N,O) as specified in Fig. 2 in the main text, that were included in the training set artificially.}
    \begin{tabular}{c|ccccc}
         X-H: & 1 [{\AA}] &2 [{\AA}] &3 [{\AA}]&4 [{\AA}]&5 [{\AA}] \\
         $\overline{\# Data point}$ &&&&&\\
         \hline
         1& 1.76	& 3.47	& 3.47	& 2.19	& 1.66 \\
2 & 1.96	& 3.57	& 3.57	& 2.37	& 1.70 \\
3& 2.06	& 3.67	& 3.67	& 2.55	& 1.76 \\
4& 2.16	& 3.77	& 3.77	& 2.73	& 1.84 \\
5& 2.26	& 3.87	& 3.87	& 2.91	& 1.89 \\
6& 2.35	& 3.97	& 3.97	& 3.09	& 2.06 \\
7& 2.45	& 4.06	& 4.06	& 3.36	& 2.11 \\
8& 2.55	& 4.56	& 4.56	& 3.45	& 2.37 \\
9& 2.65	& 5.05	& 5.05	& 3.72	& 2.43 \\
10& 2.75	& 5.55	& 5.55	& 4.07	& 2.49 \\
11& 2.85	& 6.04	& 6.04	& 4.34	& 13.98 \\
12& 2.95	& 7.03	& 7.03	& 24.88	& 22.64 \\
13& 3.00	& 8.02	& 8.02	& 28.14	& 26.59 \\
14& 3.50	& 9.01	& 9.01	& 32.10	& 32.72 \\ 
15& 3.94	& 10.00	& 10.00	& 38.44	& 44.49\\ 
16& 5.92	& 10.99	& 10.99	& 49.52	& 54.58\\
17& 10.87	& 20.88	& 20.88	& 60.31	& 62.60\\
18& 20.77	& 40.66	& 40.66	& 68.72	& 76.16\\
19& 40.58	& 80.22	& 80.22	& 75.26	& 79.13\\
20& 80.19	& 90.11	& 90.11	& 84.06	& 94.77\\
21& 100.00	& 100.00	& 100.00	& 99.90	& 97.05 \\
    \end{tabular}
\end{table}{}

\begin{table}[h]
    \centering
    \begin{tabular}{c|cc}
         X-H & r$_{eq}$ [{\AA}] &$r_{max}^{\text{ADC(2)}}$ [{\AA}] \\
         \hline
         \hline
         1 & 0.97 &1.38\\
         2 & 1.1 &1.43\\
         3 &1.1 &1.43\\
         4 &0.99 &1.48\\
         5 &1.02 &1.44\\
    \end{tabular}
    \caption{The maximum distances for ADC(2)-data usage, r$_{max}^{\text{ADC(2)}}$, and the respective distances at the equilibrium geometry, r$_{eq}$, for each X-H (X=C,N,O) bond as specified in Fig.~\ref{fig:diss}.}
    \label{tab:diss}
\end{table}{}

\subsection{Photodynamics simulations}\label{sec:dyn}
The main results of this study are based on photodynamics simulations of tyrosine that were carried out with the SchNarc approach for deep-learning enhanced nonadiabatic dynamics~\cite{Westermayr2020JPCL} with Tully's fewest switching surface hopping method~\cite{Tully1990JCP,Tully1991IJQC} as implemented in the SHARC program.~\cite{Richter2011JCTC,Mai2018WCMS} The different starting geometries for the dynamics simulations were obtained from Wigner sampling~\cite{Wigner1932MM,Wigner1932PR} of the lowest-energy conformation of tyrosine. 
The photodynamics simulations were initiated in the S$_4$ state using an excitation energy window of 6.5 to 7.0\,eV, in accordance to experiment.~\cite{Iqbal2010JPCL,Iqbal2015} Default surface hopping parameters and the energy-based decoherence correction were employed, as implemented in SHARC.~\cite{Mai2018WCMS} A time step of 0.5 \,femtoseconds was used to propagate the nuclei on the potential energy surface formed by the electrons. The potential energy surfaces were obtained from the ML predictions that were trained on \emph{ab initio} electronic structure data and adapted data (as described above). The photodynamics approach SchNarc was validated with small model systems in pilot projects.~\cite{Westermayr2019CS,Westermayr2020JPCL,Westermayr2020MLST} The accuracy and usability of this method was validated for long time scale simulations,\cite{Westermayr2019CS} ultrafast and slow nonadiabatic transitions,~\cite{Westermayr2020JPCL} and the description of different spin multiplicities.~\cite{Westermayr2020JPCL,PHDTHESIS} 

The ML model that lies at the heart of SchNarc is SchNet,~\cite{Schuett2018JCP,Schuett2019JCTC} adapted by us for excited states.~\cite{Westermayr2020JPCL} A special feature of SchNarc is that it can compute the nonadiabatic couplings between electrons and nuclei via Hessians of the squared energy-difference potentials provided by the NN model. Such approximated nonadiabatic couplings were used to determine the hopping probability from one electronic state to another during the dynamics simulations. A random number between 0 and 1 was used to decide whether a hop takes place or not, \textit{i.e.}, if the random number was within the hopping probability, a hop took place. Due to the stochastic nature of this method, many trajectories have to be carried out to obtain statistical significant results. The requirement for a large number of long time-scale trajectories makes ML models especially attractive for the purpose of this study.

Although the calculation of Hessian computations to approximate nonadiabatic couplings led to generally higher computational costs compared to the prediction of nonadiabatic couplings that were initially trained, it is, to the best of our knowledge, the only way to describe ultrafast nonadiabatic transitions in tyrosine with ML. Nonadiabatic couplings are not available in the quantum chemistry codes that support ADC(2) and were available to us. Hence, nonadiabatic couplings could not simply be fitted from ADC(2) data. For approximated nonadiabatic couplings, a threshold had to be set that ensured that nonadiabatic couplings were only approximated between two potential energy surfaces that were close enough to each other. An energy gap of 0.54\,eV was selected in this work in accordance to previous work.~\cite{Westermayr2020JPCL} If the energy gap between two states was smaller than this threshold, nonadiabatic couplings were set to 0. Due to the high density and the large number of states treated in this study, a Hessian computation for approximated nonadiabatic coupling vectors was required almost every time step, which led to high computational costs. 
Due to these computational costs that could limit long time scales, the dynamics were stopped whenever the molecule reached the S$_1$ state for at least 15 \,femtoseconds or relaxed to the S$_0$ state. After that, dynamics were carried out with a Hessian computation considering only the energy gap between the S$_0$ and the S$_1$ states, $\Delta E_{01}$. The energy gap threshold was further reduced to 0.2\,eV. 
To validate the reduction of the energy gap, the energy gaps between the S$_0$ and the S$_1$ states were predicted for the whole training set and an additionally computed geometry close to a conical intersection with CASSCF(12,11). A reasonable amount of data points was within 0.2\,eV. The reduction of the energy gap was further validated by comparing trajectories with different energy thresholds that all resulted in the same final states after 1 ps and in comparable geometrical distributions. Moreover, we propagated about 20\% of all trajectories with the initial settings, \textit{i.e.}, the initial energy gap with a Hessian calculation at almost every time step. The resulting populations were comparable to those obtained with the applied simplifications and the statistics were not altered.

As already mentioned, the Hessian calculations are quite expensive compared to the force calculations. One time step of the dynamics using the ML predictions of energies and forces for 13 states of tyrosine took up to 3 seconds on a Tesla-V100GPU and about 4-8 seconds on a 2x Intel Xeon E5-2650 v3 CPU. In contrast, a time step including the calculation of approximated nonadiabatic coupling values took about 50 seconds on the same GPU and on average about 1,000 seconds on the same CPU. Hence, the dynamics simulations of 1\,picosecond (2,000 time steps) including a Hessian calculation each time step took about 1 day on a GPU and about 23 days on a CPU.

In this work, SchNarc was extended such that multiple ML models could be used during the dynamics simulations. In total, three different ML models were executed for the dynamics. Two models were trained on energies and forces and a third one on SOCs. The ML model architecture is explained in more details in section~\ref{sec:ml}. One ML model was applied directly in the dynamics to predict energies, forces, and approximated nonadiabatic coupling vectors. Another ML model allowed for the detection of underrepresented or new conformational regions by comparison of the energy values of the two NNs. A threshold of 0.5\,eV was set for the mean absolute error (MAE) of the two NNs for the ground state energy of tyrosine, that was used to stop the dynamics whenever exceeded. To enable longer time scales, an additional adaptive sampling step was conducted using 200 trajectories. Whenever all trajectories were stopped by the adaptive sampling procedure (the longest trajectory terminated shortly before 3\,picoseconds), every fifth geometry of the 100 last time steps from each terminated trajectory was used to adapt the training set. Data points with X-H distances longer than r$_{max}^{ADC(2)}$ were set back to r$_{max}^{ADC(2)}$. In addition, we encoded in the NN models that the potentials should be constant whenever a distance between a hydrogen atom and another atom was larger than 2{\AA}. This condition was implemented to additionally ensure accurate potentials at long interatomic distances. As expected, the additional data points differed strongly from the equilibrium reference geometry, making conventional phase correction practically infeasible (more than 100s of singlets and triplet states and many interpolation steps between the equilibrium and the current geometry would have been required). To circumvent this problem, we used the NN models to find the most suitable phase vector. The NN models predicted the SOCs of every new data point. The predicted SOCs were compared to the ADC(2) SOCs and all possible 4,029 phase vectors were computed (2$^{N_{states}}$ with $N_{states}$ being the number of considered electronic states). The reference SOCs were multiplied with every phase vector and the minimum error was computed to the predicted SOCs. The corresponding phase vector was used to correct the couplings that were then added to the training set. This approach is similar to the phase-free training algorithm implemented in SchNarc. 
The NN models were then re-trained with the extended training set of 17,265 data points and the dynamics simulations were restarted. In addition, 844 trajectories were started from additional 6,000 Wigner sampled geometries of tyrosine that were excited to the S$_4$ state.
In this way, the simulation of thousand trajectories in the picosecond time scale could be achieved for statistical significant results, that would have taken up to 8 years for a single trajectory with CASPT2 reference calculations. A larger threshold of 1\,eV was set for the NN models used during the final dynamics to terminate the trajectories. 

\begin{figure}[ht]
    \centering
    \includegraphics[scale=0.5]{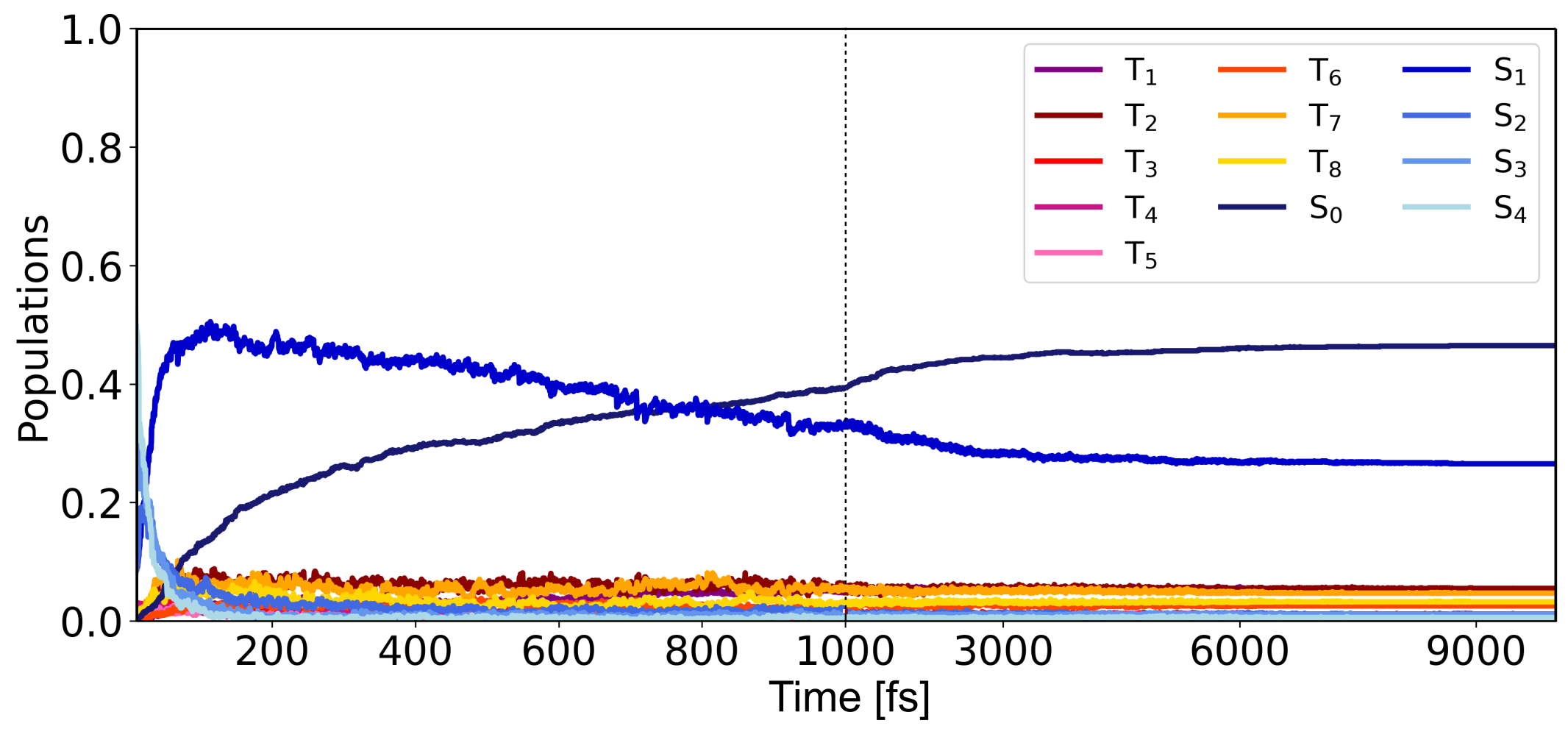}
    \caption{Population curves of five singlet and eight triplet states of tyrosine excited to the fourth excited singlet state, S$_4$, obtained from averaging 1,022 trajectories up to 1\,picosecond, 261 trajectories up to 2\,picoseconds and 127 trajectories longer than 2\,picoseconds.}
    \label{fig:pop10ps}
\end{figure}

The population curves up to 10\,picoseconds in addition to Fig. 4(a) in the main text are shown in Fig.~\ref{fig:pop10ps}.

\subsection{Validation of roaming}
\begin{figure}[ht]
    \centering
    \includegraphics[scale=0.33]{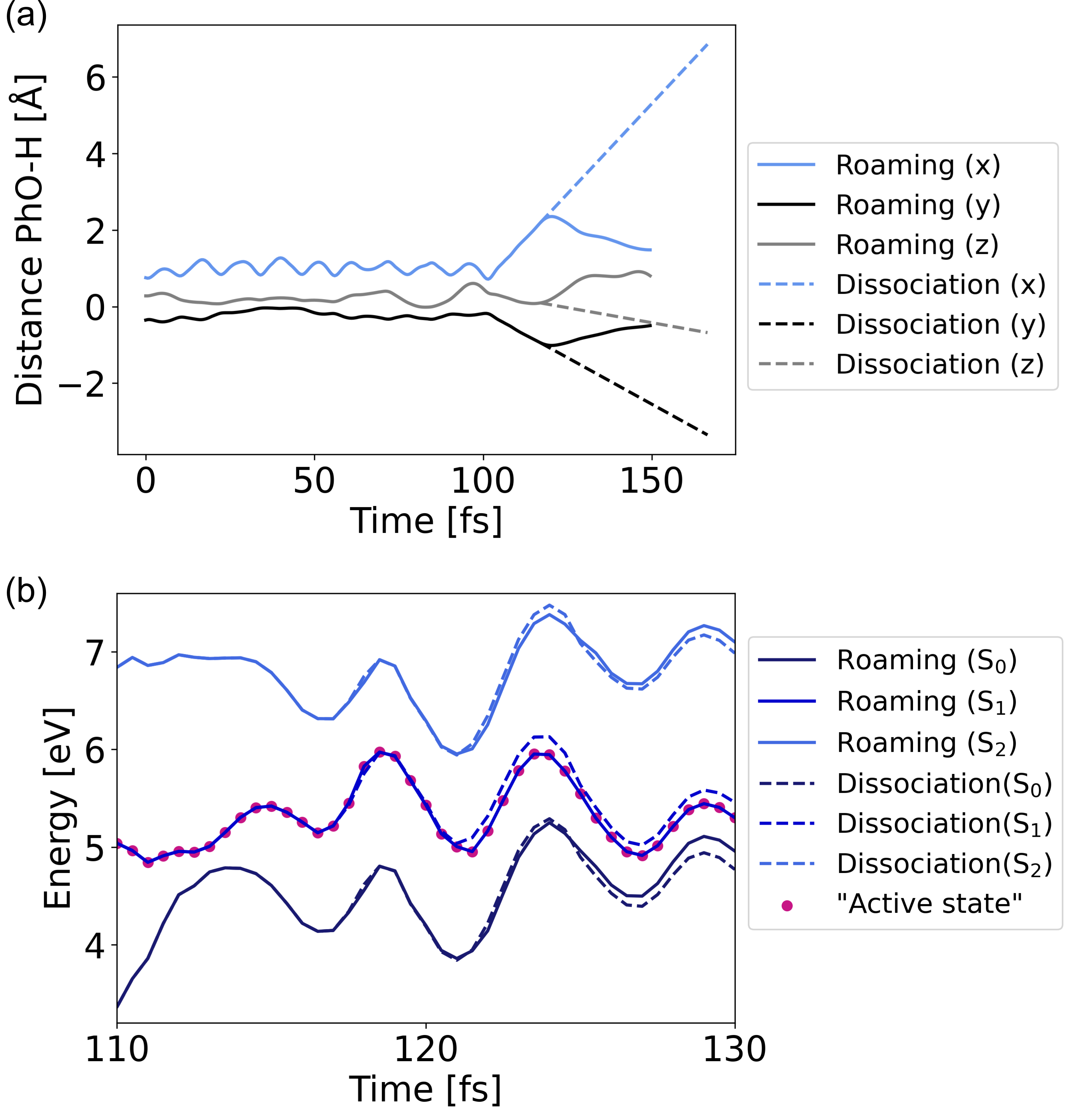}
    \caption{(a) The bond distance of the H-atom attached to the PhO-H group along the simulation time. The distances of the trajectory that mimicks dissociation is shown by dashed lines. (b) The potential energy curves of the original and the mimicked dissociative trajectory are computed with CASPT2/ano-rcc-pVDZ. The active state at a given time step is marked with red dots.}
    \label{fig:roaming}
\end{figure}

The event of roaming atoms was verified using CASPT2(12,11)/ano-rcc-pVDZ reference calculations and an example trajectory that showed a roaming atom between 100\,femtoseconds and 150\,femtoseconds. The bond distances of the hydrogen atom bound to the PhO-H group in x, y, and z directions are plotted along the simulation time in Fig.~\ref{fig:roaming}. As can be seen, after about 100\,femtoseconds of simulation time, the bond distance starts to increase in the x and z directions. At a time of about 120\,femtoseconds, the H atom changes direction and the distance shortens again. This is the time at which roaming was favored compared to the conventional dissociation pathway, which is shown by dashed lines. We mimicked the dissociation pathway by extending the PhO-H bond in the direction illustrated by dashed lines, \textit{i.e.}, the direction in which the H atom was moving before the change in direction occurred. The rest of the atoms remained unchanged from the original trajectory. The potential energy curves of the first three singlet states of both trajectories are shown in panel (b). The active state of the molecule is the S$_1$ state, which is marked by red dots. As can be seen, the roaming pathway is energetically favored compared to the dissociative pathway, confirming the likely nature of roaming. The molecular structures that entered the analysis are attached as xyz-files in the supplementary material.

\begin{figure}[ht]
    \centering
    \includegraphics[scale=0.6]{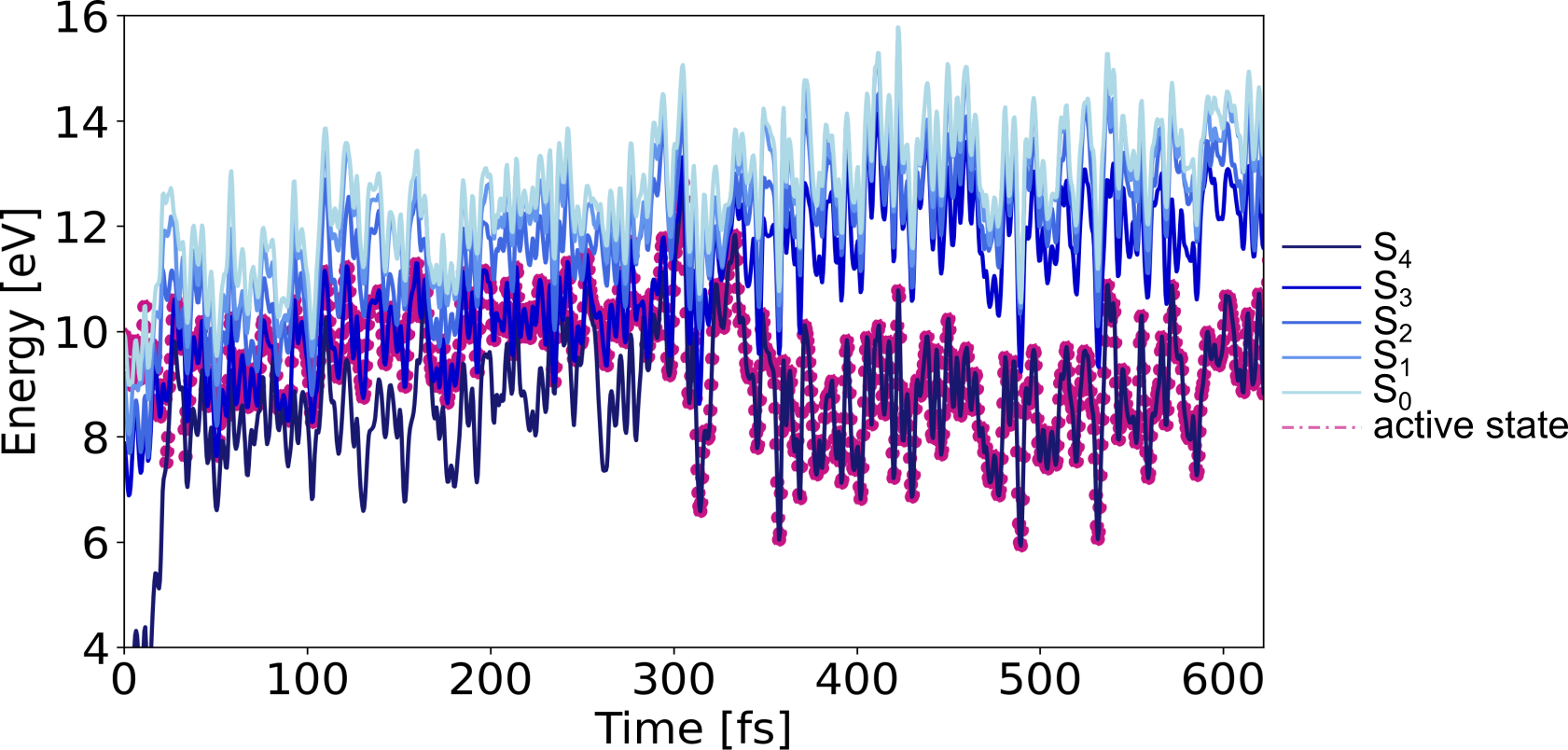}
    \caption{Potential energy curves of five singlet states of the trajectory shown in Fig. 1 in the main text. The active state is marked at each time step. For better visibility of the separate states, the energy is shown within an range of 4 to 16\,eV. }
    \label{fig:mainfig1}
\end{figure}

The potential energy curves that correspond to Fig. 1 in the main text are shown in Fig.~\ref{fig:mainfig1} with the active state indicated at every time step.

\subsection{Analysis of dynamics}
We applied an enlarged threshold of 1\,eV between two NN models to let dynamics run longer than they would if adaptive sampling was applied. Hence, analysis of dynamics was carried out up to a time step in which the total energy did not rise by more than 5\,eV. A constant, slight rise in the energy is expected for longer dynamics.~\cite{Westermayr2019CS} The longest trajectory reached 13,118\,femtoseconds before previously unexplored conformations were encountered. 22 trajectories were sorted out from the 1,044 (200 + 844) trajectories after analysis, resulting in 1,022 trajectories in total. Reasons for sorting out trajectories were for instance a non-converged singular value decomposition during the dynamics or large jumps in the total energy in consecutive time steps. 

To split the trajectories into roaming and non-roaming, the PhO-H bond length was analyzed along the trajectories. If the PhO-H bond length was larger than 2.5 {\AA} and the minimum distance of the H atom to another atom in the molecule was smaller than 5 {\AA}, the trajectory was classified as roaming. We additionally analyzed N-H, COO-H, and C-H bond lengths, but those did not change the classification of trajectories into roaming or non-roaming. A trajectory was classified into dissociation, when the distance of an H atom to any other atom in the parent molecule exceeded 5 {\AA}. Each following analysis was carried out for a set of trajectories independently. Four sets, (a) no roaming without dissociation, (b) no roaming with dissociation, (c) roaming without dissociation, and (d) roaming with dissociation were used.

\subsubsection{Geometrical analysis: Clustering}
\begin{figure}[ht]
    \centering
    \includegraphics[scale=0.45]{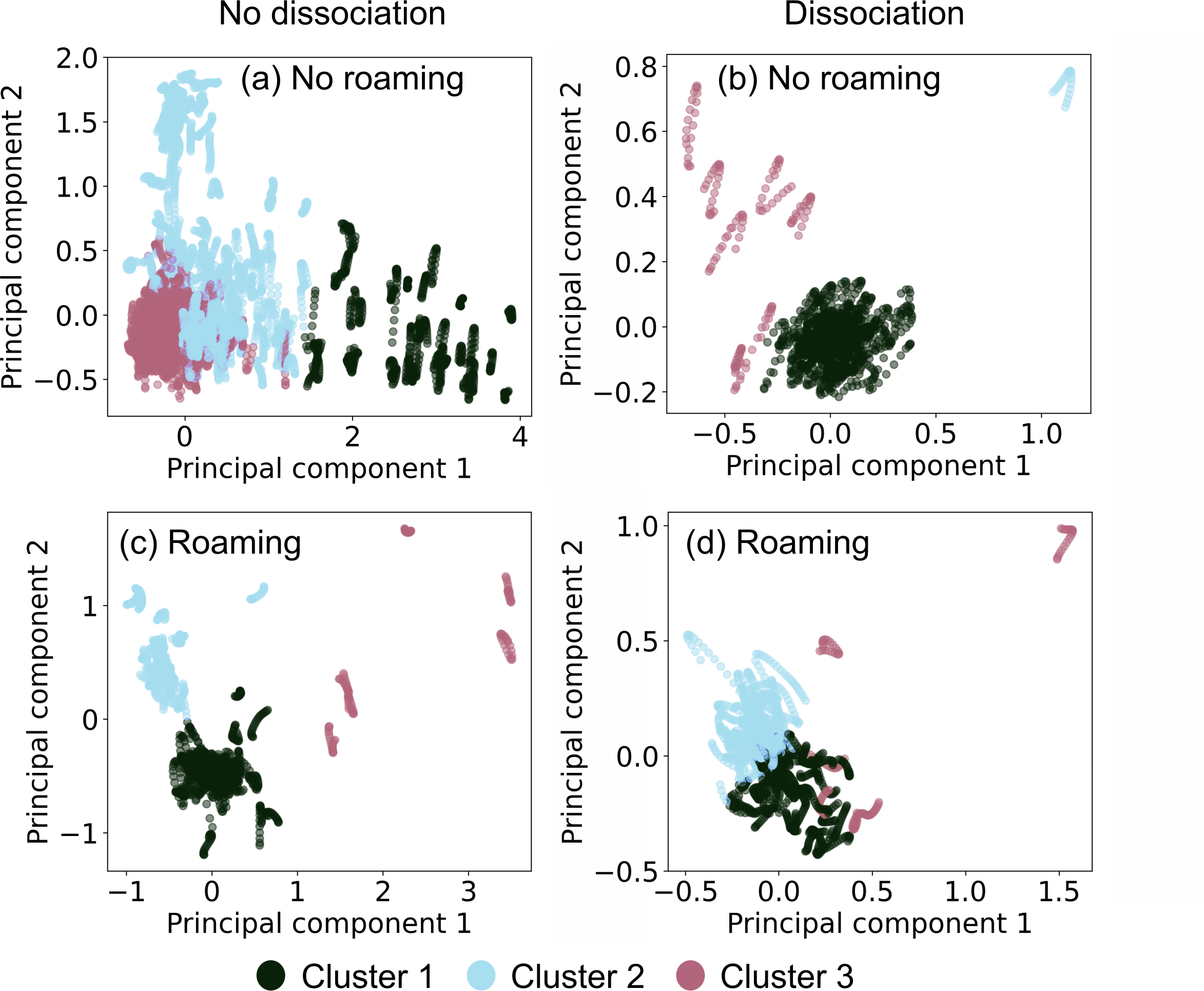}
    \caption{Three initial clusters of products of (a-b) non roaming and (c-d) roaming trajectories that were additionally separated into dissociation (right block) and no dissociation (left block). Each cluster was additionally clustered into 4 to 6 clusters.}
    \label{fig:cluster}
\end{figure}

The geometrical analysis shown in Fig. 2 in the main text was carried out using k-means clustering as implemented in scikit-learn.~\cite{scikit-learn} Geometries of the last 10\,femtoseconds of each trajectory were used for analysis.
Non-roaming and roaming trajectories were treated separately and each set was further split into trajectories that showed dissociation and those that did not show dissociation. 
K-means clustering required a molecular descriptor, which we chose to be the inverse distance matrix (without contributions from the dissociated atom in case of dissociation trajectories), and a number of clusters as an input. The initial number of clusters was set to 3. In the context of this work, principal component analysis was used to reduce the dimensions of the descriptor, which is a 24$\times$24 matrix, and to visualize the distribution of the data points along the first two principal components that cover most variation in the data. Principal component analysis was carried out with tools implemented in scikit-learn. We note that this method could be applied before clustering as well, which we tested with a set of principal components that covered up to 90\% of the variance in the data. This procedure led to complementary results, hence principal component analysis was used solely for visualization in this work.
The distribution of the data along the principal components for each set (a)-(d) is shown in Fig.~\ref{fig:cluster}. The three initial clusters were additionally clustered in a second round into 4 to 6 clusters to ensure all different types of products and rare events were recovered in the analysis. 

\subsubsection{Charge analysis}
To find out whether the bond between the roaming atom and the parent oxygen atom is homo- or heterolytically broken, leading to radicals or cations and anions, respectively, the partial atomic charges were analyzed.
The partial atomic charges were predicted by another deep neural network model (see section~\ref{sec:dipoles} for details) trained on dipole moment vectors. All molecular conformations defined as roaming in the previous section were used for the analysis. The mean partial charge of roaming atoms in 10,767 molecular conformations was 0.019 $\pm$ 0.038 and that of parent oxygen atoms was 0.083 $\pm$ 0.026. The maximum/minimum value of partial charges found in roaming atoms and parent oxygen atoms were 0.16/-0.13 and 0.24/-0.11, respectively. The atomic partial charges of the active state the molecule was in during the dynamics was used. 
The results strongly suggest that roaming atoms are present as radicals rather than cations.

\subsubsection{Kinetics}

\begin{figure*}[ht]
    \centering
    \includegraphics[scale=0.8]{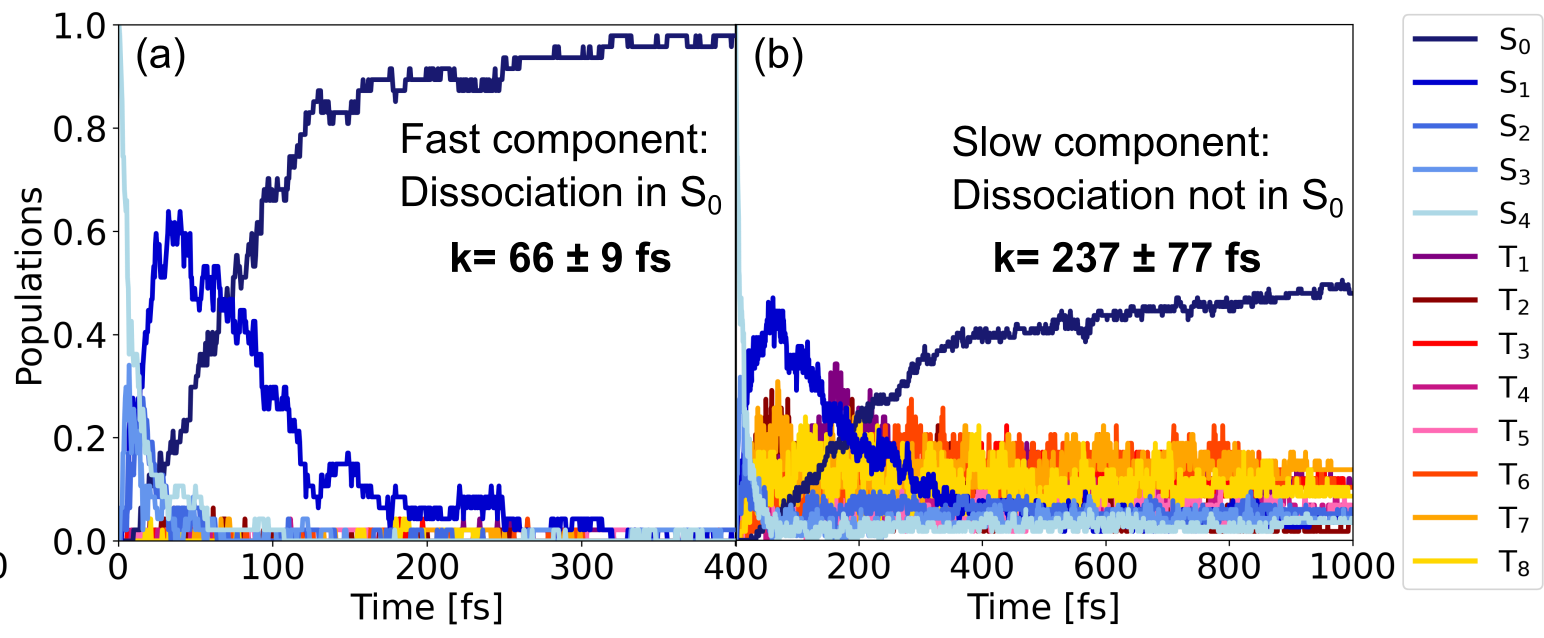}
    \caption{Population plots of trajectories that show dissociation of a hydrogen atom (a) in the ground state, S$_0$, (47 trajectories) and (b) in an excited state (117 trajectories). Fitting of the time constants was carried out using the population curve of the S$_0$ state and the tools available in SHARC.~\cite{Mai2018WCMS,sharc-md2}.}
    \label{fig:kinetic}
\end{figure*}

The kinetics of the dissociation was analyzed by splitting the trajectories into a set that dissociated in the ground state and a set that dissociated in an excited state. Dissociation was defined to take place when the bond-distance of an H atom to the tyrosine parent fragment exceeded 5 {\AA} and was not roaming around the molecule. The population plots obtained from each set of trajectories are shown in Fig.~\ref{fig:kinetic}(a) and (b). To obtain time constants of the reaction, the ground state poulation was fitted using the analysis tools of SHARC.~\cite{Mai2018WCMS,sharc-md2} We obtained a time constant of k$_1$ = 66 $\pm$ 9\,femtoseconds and of k$_2$ = 237 $\pm$ 77\,femtoseconds for the fast and slow component, respectively.
The analysis is complementary to Fig. 3(e), in which the amount of dissociation occurring in all trajectories is shown along the simulation time. This curve could be fit with an exponential function, $y=a_1 \exp{-\frac{t}{k_1}}+a_2\exp{-\frac{t}{k_2}}$, using the fitted time constants from the population curves, and thus verifying the assumption of experimental studies that the two time constants result from dissociation in different electronic states.

\subsection{Machine learning potentials}\label{sec:ml}
In this work, ML models were trained on a combined data set computed with ADC(2) and augmented to represent CASPT2(12,11) potential energy curves at long interatomic distances. It is worth mentioning that other possibilities exist to augment an ML model with higher accurate data. For instance, ML models could be trained on a cheap data set and adapted to reproduce values from a more accurate method by $\Delta$-learning approaches.~\cite{Ramakrishnan2015JCTC,Zaspel2019JCTC,Westermayr2021arXiv} One model is trained on many points from a cheap data set and the other model on a few points of the energy difference from a cheap data set and an expensive data set. This approach, however, requires either two inferences or a cheap reference, baseline method. In the case of tyrosine, this approach was expected to yield inaccurate results due to the inconsistencies in the expensive method (CASPT2) along the potential energy curves (see Fig.~\ref{fig:methods}(a)). Moreover, this approach needs additional considerations and adaptions in the case of couplings and has not been tested for other purposes than potential energy predictions. In contrast, transfer learning exists that requires one inference, but two training steps. In this approach, a model is first trained on a cheap data set and the parameters of this model are fixed except for a few nodes. Afterwards, this rigid model is retrained on a second, more accurate training set. This procedure was carried out for example for the ANI-1ccx NN.~\cite{Smith2019NC} The inconsistent potential energy curves of CASPT2 are also considered problematic for the application of transfer learning.
\subsubsection{Initial training of multi-layer feed-forward NNs}
Multi-layer feed-forward NNs with the inverse distance matrix as a descriptor for molecular structures were used to generate the initial training set for photodynamics simulations.
\subsubsection*{Hyperparameters}
The hyperparameters of two distinct NNs are given in Table~\ref{tab:hyperparams} for the training set of 16k data points with and without artificial data points. The forces were treated as derivatives of the potentials for energies as it was described in our previous publication.~\cite{Westermayr2019CS} The mean absolute error (MAE) of energies, forces, and SOCs on the test set is 0.0227\,eV, 1.25\,eV/{\AA}, and 0.0971\,cm$^{-1}$, respectively, for the model with the final training set including artificial data points. 
\begin{table*}[ht]
\caption{\label{tab:hyperparams}Parameters for different NN models (NN1 and NN2). The inverse distance matrix was used as a molecular representation. E, F, and SOCs is the abbreviation for energies, forces, and SOCs, respectively. The batch size was set to 500 with a maximum of 5000 epochs carried out. The number of hidden layers was set to 8 for energies and forces and to 15 for SOCs with 50 nodes per hidden layer. As a basis function, the shifted softplus function, $ln(0.5 e^{x}+0.5)$, and the hyperbolic tangent were used for energies+forces and SOCs, respectively. After a given number of steps ("Steps for annealing $lr$"), the learning rate, $lr$, was annealed by multiplication with a factor ("Factor to anneal $lr$"). }\begin{tabular}{l|llll}
\hline
\hline
 &NN1 (E+F) &NN1 (SOCs) &NN2 (E+F) &NN2 (SOCs)\\
\hline 
Learning rate, $lr$ &$6.4\cdot 10^{-4}$     &$7.8\cdot 10^{-4}$     &$4.7\cdot 10^{-4}$     &$9.1 \cdot 10^{-4}$ \\
Factor to anneal $lr$ &0.9767 &0.9795   &0.9875 &0.9633\\
Steps for annealing $lr$ &69 &70 &672 &49\\
L2 regularization &1.0 $\cdot 10^{-9}$ &1.0 $\cdot 10^{-9}$ &1.0 $\cdot 10^{-9}$ &1.0 $\cdot 10^{-9}$  \\
\end{tabular}
\centering
\end{table*}
\subsubsection*{Learning curves}
\begin{figure*}[ht]
    \centering
    \includegraphics[scale=0.28]{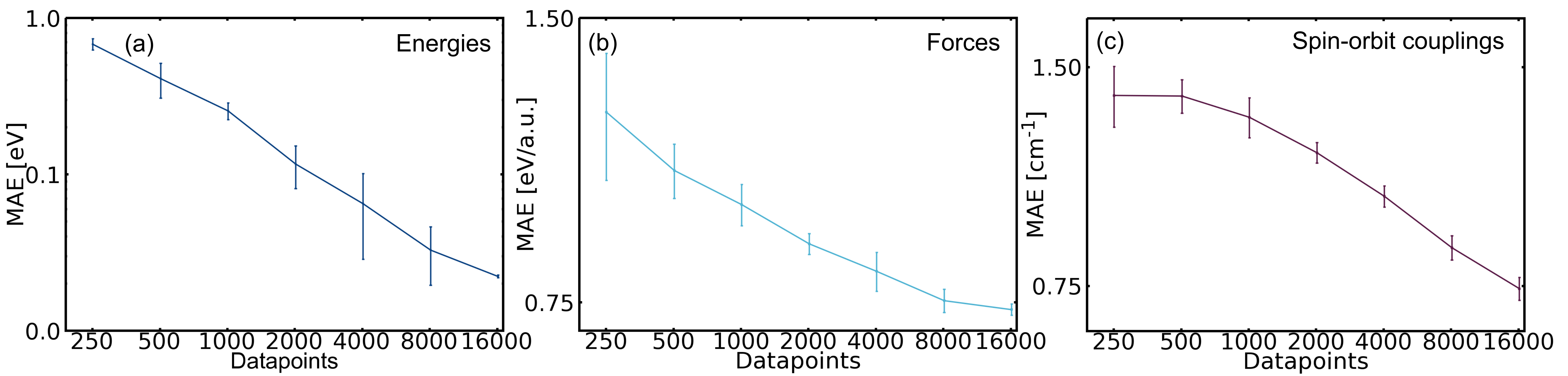}
    \caption{Learning curves showing the mean absolute error (MAE) for the (a) energies, (b) forces, and (c) SOCs averaged over all states and pairs of states using the final training set including artificially generated data points.}
    \label{fig:lcfinal}
\end{figure*}

\begin{figure}[ht]
    \centering
    \includegraphics[scale=0.1]{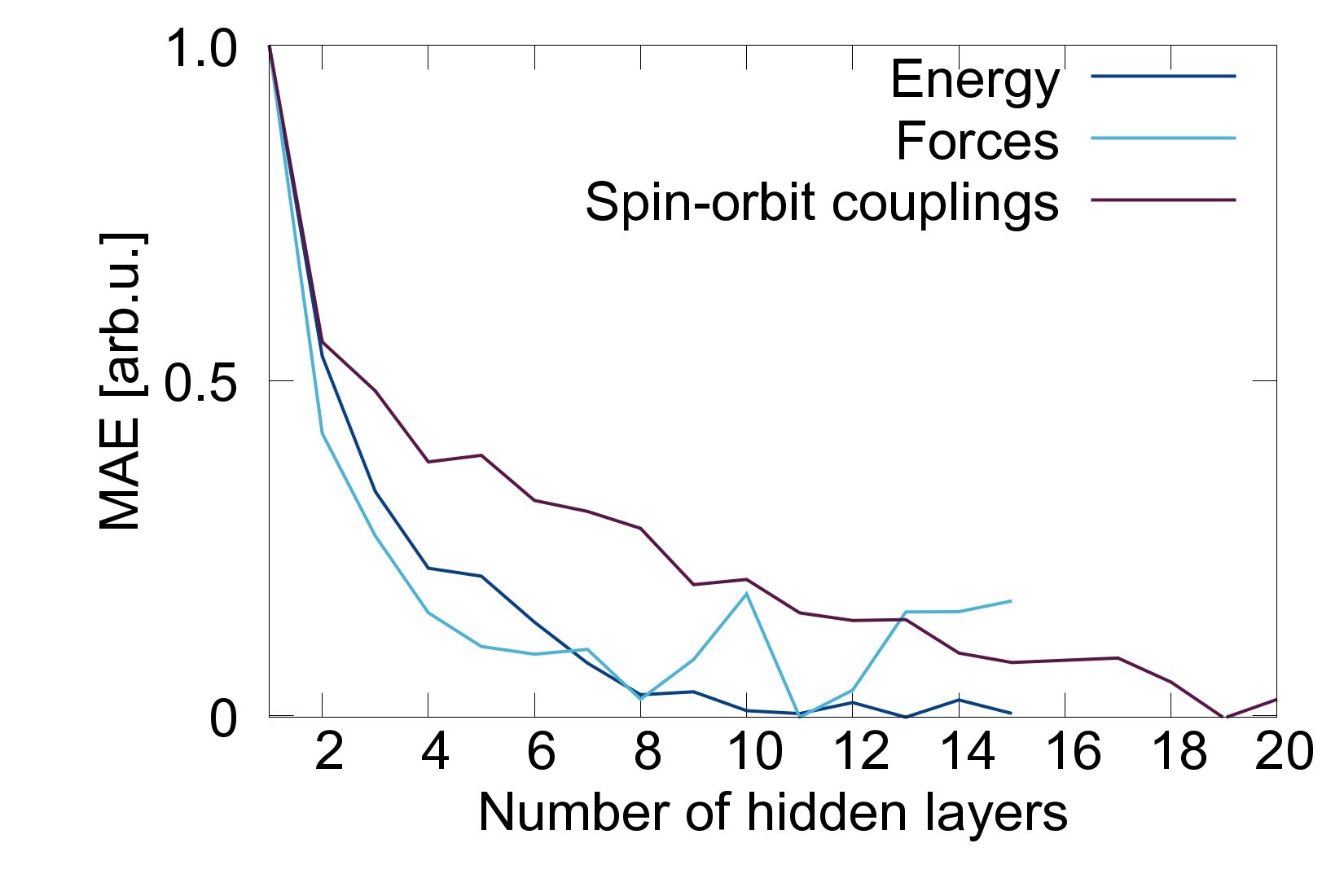}
    \caption{The mean absolute error (MAE) plotted in arbitrary numbers against the number of used hidden layers for the neural network architecture for energies, forces, and SOCs of the final training set.}
    \label{fig:hl}
\end{figure}

The learning curves of the models trained on the training set including artificially generated data points is shown in Fig.~\ref{fig:lcfinal}. The MAEs plotted against the number of hidden layers used for a network architecture with 50 nodes per hidden layer is shown in Figure ~\ref{fig:hl}.
The number of hidden layers used for energies and forces was set to 8, whereas 15 was assumed to be sufficient for SOCs. Learning curves were computed to evaluate whether the model could learn from these data or not. As can be seen, the error decreased with increasing training set size, hence the training set was used for the training of the final SchNet NN model.
{}
{}

\subsubsection{ML models for photodynamics: The SchNarc approach}
\begin{figure*}[ht]
\centering
\includegraphics[scale=0.8]{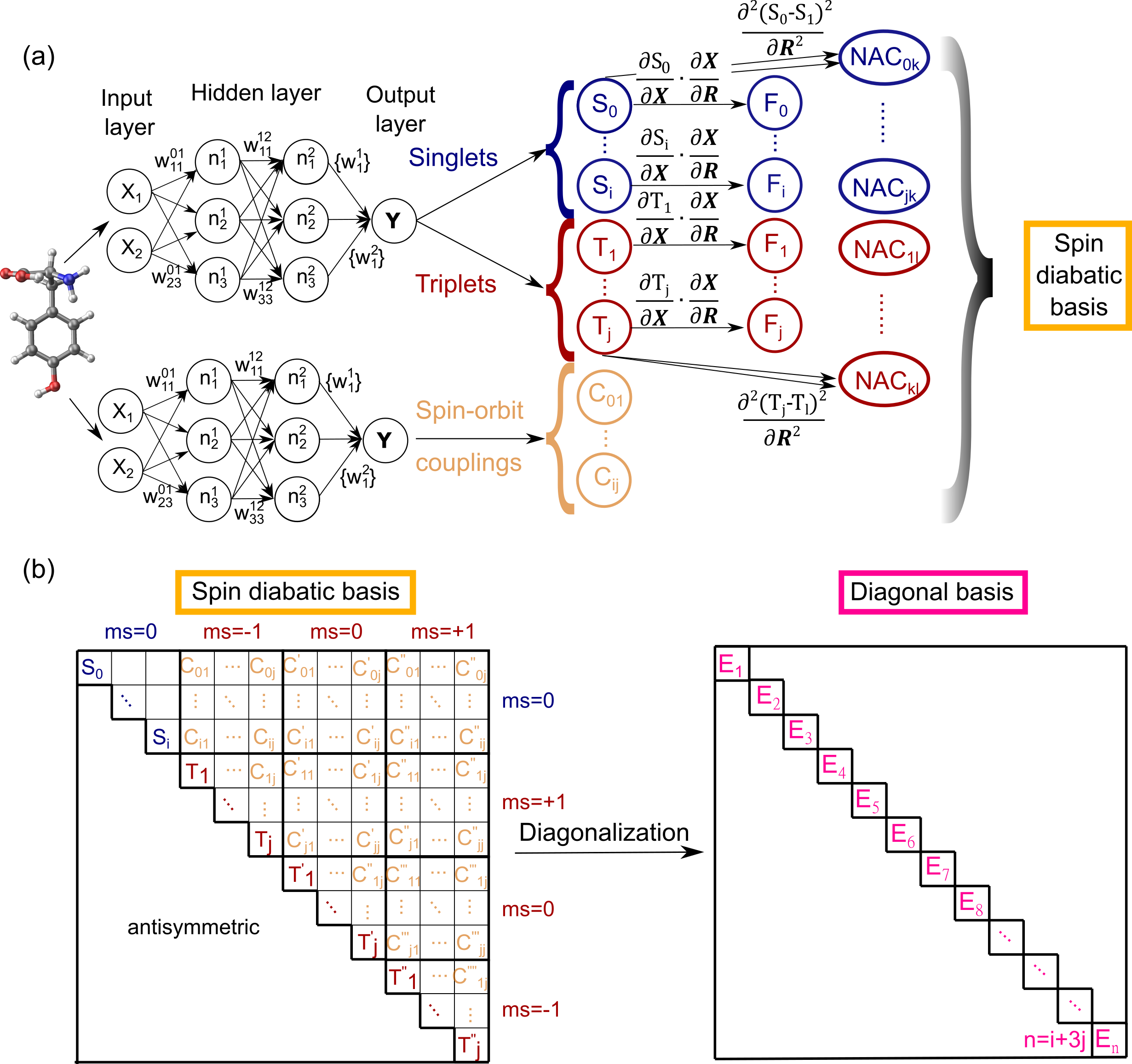}
    \caption{(a) A schematic representation of the two NNs that were used during the dynamics simulations to propagate the nuclei of the molecule. One NN was used to predict energies, forces, and approximated nonadiabatic coupling vectors, while the other was used for prediction of SOCs. The outputs of both NNs were used to fill a spin diabatic Hamiltonian matrix ("molecular Coulomb Hamiltonian" (MCH) basis in SHARC~\cite{Mai2018WCMS,Mai2015IJQC}). (b) The diagonal Hamiltonian matrix (also known as fully adiabatic) is obtained after diagonalization of the spin diabatic matrix. }
    \label{fig:ml}
\end{figure*}
The final model used for production runs with SchNarc was trained using the message-passing continuous filter convolutional NN SchNet for excited states.~\cite{Schuett2018JCP,Schuett2019JCTC,Westermayr2020JPCL} 
The structure of the NNs that were used for photodynamics simulations can be seen in Fig.~\ref{fig:ml}(a). 
The hyperparameters of the NN models were sampled on a random grid. This procedure was sufficient as the descriptor of the molecular structure was additionally learned and thus tuned to represent the data in the most optimal way. We sampled the learning rate, number of features, layers, and Gaussian functions for the representation block of SchNet that fits the molecular descriptor and the number of hidden layers and nodes in the prediction block of SchNet that relates the tailored descriptor to the targeted property. In addition, different learning rates, batch sizes, and cutoffs were evaluated. The final models were trained on 15,500 data points with 500 data points for validation to avoid overfitting. The remaining data points were used for testing (8,502 values for energies and forces and 531,048 SOC values). The split was done randomly by SchNet. For training, a cutoff of 6.0\,a.u. was used to model an atom in its chemical and structural environment. A batch size of 16 was used in combination with 512 features, 4 interaction layers, and 25 Gaussian functions for the representation block and 4 hidden layers and 100 nodes per hidden layer for the prediction block. The remaining hyperparameters remained default values of SchNet. The SOC model was trained using 256 features, 3 hidden layers and 128 nodes per hidden layer with a batch size of 32.

As can be seen in panel Fig.~\ref{fig:ml}(a), the molecular structure was given to the NN model, which learned the representation of the molecule in addition to the targeted outputs. For the first ML model, the targeted outputs were the singlet and triplet potential energy surfaces, $S_i$ and $T_j$, respectively, including their derivatives, $\mathbf{F}_i$ and $\mathbf{F}_j$, which were derived with respect to the molecular descriptor, $\mathbf{X}$, and multiplied with the partial derivative of $\mathbf{X}$ with respect to the Cartesian coordinates of the molecule, $\mathbf{R}$. The second derivative of the squared energy difference potentials could be used in combination with gradient difference vectors to approximate nonadiabatic coupling values. As can be seen, the SOCs were treated in a separate NN model. All properties were in the spin-diabatic basis~\cite{Mai2015IJQC} (known as the "molecular Coulomb Hamiltonian" (MCH) basis in SHARC). 

The spin-diabatic Hamiltonian is shown in more details in panel Fig.~\ref{fig:ml}(b) on the left side. As can be seen, singlet and triplet states are separated and SOCs couple the different states. The spin-diabatic Hamiltonian is the direct output of any quantum chemistry calculation. In contrast, a diagonal (or spin-adiabatic) Hamiltonian contains spin-mixed states that are split by the relevant SOCs. 

The ML model for energies and forces used a combined loss function with manually defined trade-offs for energies and forces, $t_E$ and $t_F$, which were set to 1 and 0.1, respectively.
The loss function, $L_{EF}$, that was optimized during training thus reads:
\begin{eqnarray}
    L_{EF} = t_E \cdot \left[ \frac{1}{N_S}\sum_i^{N_S}\left( S_i^{ML}- S_i^{Ref.}\right)^2 \color{white}\right]  \\
\nonumber   \color{white}\left[ \color{black} + \frac{1}{N_T}\sum_j^{N_T}\left( T_j^{ML}- T_j^{Ref.}\right)^2 \right] \\
        \nonumber + t_F \cdot \left[ \frac{1}{N_S\cdot N_A}\sum_i^{N_S} \sum_a^{N_A} \left(\frac {\partial S_i^{ML}}{\partial \mathbf{R_a}}- \mathbf{F}_{ai}^{Ref.}\right)^2 \color{white}\right]  \\
    \nonumber  \color{white}\left[ \color{black} +\frac{1}{N_T\cdot N_A}\sum_j^{N_T} \sum_a^{N_A} \left(\frac {\partial T_j^{ML}}{\partial \mathbf{R_a}}- \mathbf{F}_{aj}^{Ref.}\right)^2 \right] . \nonumber  
\end{eqnarray}
$N_S, N_T$, and $N_A$ are the number of singlet states, triplet states, and atoms, respectively. $S_i^{ML}$/$T_j^{ML}$ and $S_i^{Ref.}$/$T_j^{Ref.}$ refer to the predicted singlet/triplet states by the ML model and the reference method, respectively. $\mathbf{F}_{ai}/\mathbf{F}_{aj}$ refer to the force vector of atom $a$ in Cartesian coordinates (x,y,z) for a singlet/triplet state.

For training of SOCs we found that SOCs could be learned more accurately when they were trained in combination with the diagonal energies, $E_n$, that were obtained after diagonalization of the full spin-diabatic Hamiltonian. As can be seen in panel Fig.~\ref{fig:ml}(b), the spin-diabatic Hamiltonian was obtained as a combination of SOCs and spin-diabatic energies, $S_i$ and $T_j$, that had to be learned in addition. 
The loss function, $L_{C}$, for learning SOCs was a combination of energies, forces, and SOCs with another trade-off value for SOCs, $t_{C}$, which was set to 500 in this work:
\begin{eqnarray}
    L_C = L_{EF}  \\
    \nonumber + t_{C} \left \{ t_{c2} \left[ \frac{1}{N_C} \sum_{i,j>i}^{N_C} \left( C_{ij}^{ML} - C_{ij}^{Ref.}\right)^2 \right] \color{white}\right\} \\
     \color{white}\left\{ \color{black} 
    \nonumber + t_{C} \left[ \frac{1}{N_S+3N_T}\sum_n^{N_S+3N_T} \left( E_n^{ML}-E_n^{Ref.}\right)^2 \right] \right\}.
\end{eqnarray}
$C_{ij}^{ML}$ and $C_{ij}^{Ref.}$ refer to the SOCs predicted by the ML model and the reference SOCs, respectively, with $N_C$ referring to the number of coupling values, $N_C = \left[ \left( N_S+3 \cdot N_T\right)^2 -\left(N_S+3 \cdot N_T \right) \right] $. 
The diagonal energies, $E_n$, were are obtained as the eigenvalues of the Hamiltonian in the diagonal basis, $\mathbf{H}^{diag}$, which was obtained via diagonalization of the spin-diabatic Hamiltonian, $\mathbf{H}^{spin-diabatic}$:~\cite{Mai2018WCMS,Mai2015IJQC}
\begin{equation}
    \mathbf{H}^{diag} = \mathbf{U}^\dag \mathbf{H}^{spin-diabatic}\mathbf{U}.
\end{equation}
Noticeably, the diagonal energies of the reference method were not computed every training epoch, but were saved in the data base prior to training as an additional entry next to the spin-adiabatic energies, forces, and SOCs. It is worth mentioning that the diagonal representation is used during the dynamics simulations.

\begin{figure*}[ht]
    \centering
    \includegraphics[scale=0.45]{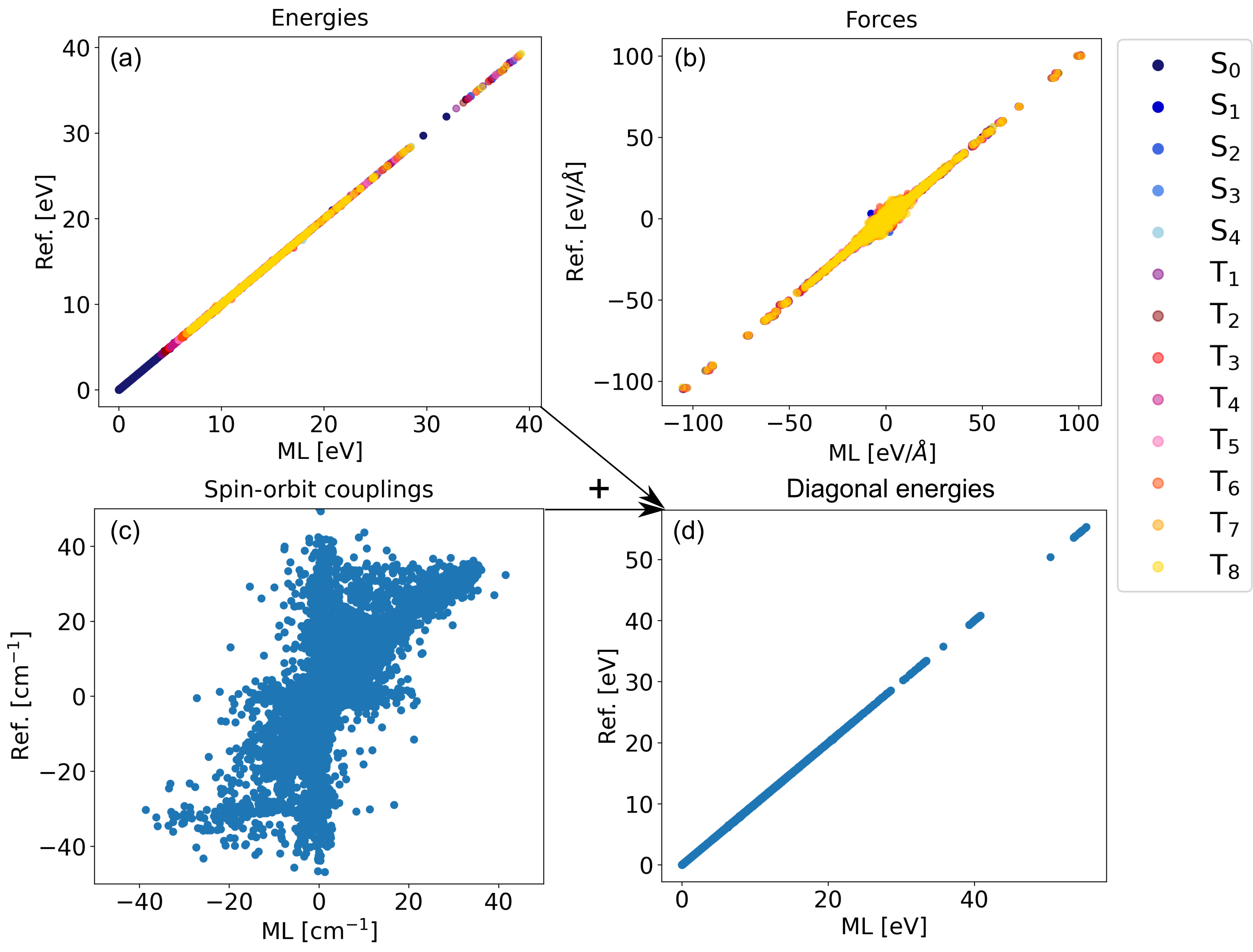}
    \caption{Scatter plots of (a) energies, (b) forces, and (c) spin-orbit couplings in the spin-diabatic basis to obtain (d) spin-adiabatic eneriges. The x-axis shows ML predictions and the y-axis refers to the reference values.}
    \label{fig:scatter}
\end{figure*}

By learning both, SOCs and diagonal energies together, the accuracy of predicted SOCs could be increased by over 22\%, allowing for accurate dynamics including relativistic effects. The final energy models were up to 10 times more accurate in terms of mean absolute errors (MAEs) for energies than previously reported models for excited states.~\cite{Westermayr2020JCP,Westermayr2020MLST,Westermayr2019CS} The MAE for energies was 9.6\,meV and for forces 163\,meV/{\AA}. 
The final SOC model yielded an MAE of 0.137\,cm$^{-1}$. Although this error was slightly larger than the error obtained with the multi-layer feed-forward NNs, the combination of both models led to highly accurate diagonal energies with an average MAE of 8.1\,meV. The MAEs of the adiabatic energies and the diabatic energies are listed state-wise in Table~\ref{tab:maes}. 
Scatter plots of energies and forces can be seen in Fig.~\ref{fig:scatter}(a) and (b), respectively. The SOCs are shown in panel (c) and the diagonal (spin-adiabatic) energies as a result of (a) and (c) are shown in panel (d).

\begin{table}[ht]
    \centering
    \begin{tabular}{c|c||c|c}
         Spin-diabatic state & MAE [meV] &Diagonal state & MAE [meV]  \\ \hline \hline 
         S$_0$& 6.3  &1 & 5.0\\
         S$_1$ & 7.5 &2 &5.5 \\
         S$_2$ & 7.5 &3 & 5.5\\
         S$_3$ & 8.6 &4& 5.5\\
         S$_4$ & 10.6 &5 &7.2 \\
         T$_1$ (ms=-1) & 6.8 &6 &7.2 \\
         T$_2$ (ms=-1) & 8.5 &7&7.2\\
         T$_3$ (ms=-1)& 10.3 &8&6.7\\
         T$_4$ (ms=-1)& 9.6 &9&8.1\\
         T$_5$ (ms=-1)& 11.2&10& 8.3\\
         T$_6$ (ms=-1)& 12.2&11& 7.5\\
         T$_7$ (ms=-1)& 12.8&12& 7.3\\
         T$_8$ (ms=-1)& 12.3&13 &7.8\\
                  T$_1$ (ms=0) & 6.8 &14 &7.9 \\
         T$_2$ (ms=0) & 8.5 &15&7.2\\
         T$_3$ (ms=0)& 10.3 &16&7.8\\
         T$_4$ (ms=0)& 9.6 &17&8.5\\
         T$_5$ (ms=0)& 11.2&18&8.4\\
         T$_6$ (ms=0)& 12.2&19&8.3\\
         T$_7$ (ms=0)& 12.8&20&9.4\\
         T$_8$ (ms=0)& 12.3&21 &9.6\\
         T$_1$ (ms=+1) & 6.8 &22 &9.5 \\
         T$_2$ (ms=+1) & 8.5 &23&9.5\\
         T$_3$ (ms=+1)& 10.3 &24&10.0\\
         T$_4$ (ms=+1)& 9.6 &25&10.0\\
         T$_5$ (ms=+1)& 11.2&26&9.2\\
         T$_6$ (ms=+1)& 12.2&27&10.0\\
         T$_7$ (ms=+1)& 12.8&28&9.9\\
         T$_8$ (ms=+1)& 12.3&29&9.7 \\
    \end{tabular}
    \caption{Accuracy of the NN model trained on spin-diabatic energies without (MAEs referring to the adiabatic states) and with the SOC ML model (MAEs referring to the diagonal states) on a test set of 654 individual data points, \textit{i.e.} 8,502 spin-diabatic energies, 18,966 spin-diabatic energies and 531,048 SOCs. }
    \label{tab:maes}
\end{table}

\subsubsection{Charge analysis}\label{sec:dipoles}
\begin{figure}[ht]
    \centering
    \includegraphics[scale=0.9]{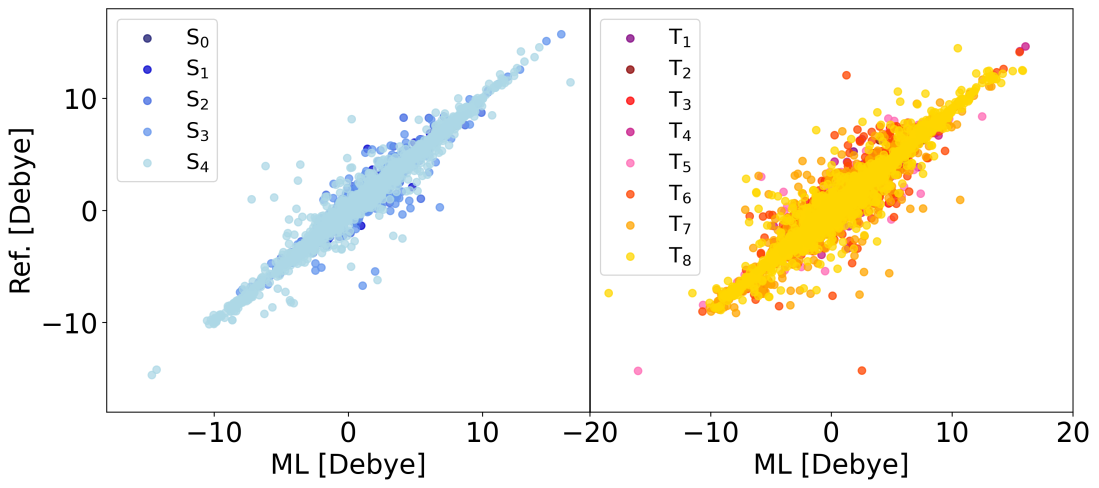}
    \caption{Scatter plots of the predicted permanent dipole moments plotted against the reference permanent dipole moments for singlet (left panel) and triplet states (right panel).}
    \label{fig:dipole}
\end{figure}

In ref. \citenum{Westermayr2020JCP}, we extended the SchNarc approach to fit dipole moment vectors, $\vec{\mu}$, via latent partial atomic charges for a given state i, $q_{i,a}$. This model was used here to analyze the atomic partial charges of the roaming atoms and the parent oxygen atoms. It is worth mentioning that partial charges cannot be obtained by solving the electronic Schr\"odinger equation and that their computation requires a post-processing step. Different charge models exist, such as the Hirshfeld~\cite{Hirshfeld1977TCA} or Mulliken partitioning scheme.~\cite{Mulliken1955JCP} However, especially the latter scheme, is often considered as unreliable and less accurate than the former. Further, these schemes are often not implemented for the excited states.~\cite{Gastegger2017CS,Westermayr2020JCP,Gastegger2020} Nevertheless, the models can be validated with dipole moment vectors for the ground and excited states. The dipole moment vectors are obtained as the sum of partial atomic charges multiplied by the vector that points from an atom to the center of mass:
\begin{equation}\label{eq:mu}
    \vec{\mu}_{i} = \sum_a^{N_a} q_{i,a} r^{CM}_a
\end{equation}
The model fits the permanent and transition dipole moments. The latter properties are fitted in the same way as permanent dipole moment vectors to preserve rotational covariance. 
By introducing physics into the models, the charge distribution in a molecular system can be accurately described. Therefore, a deep NN was trained on permanent and transition dipole moments of the phase corrected training set in this study. 
The MAE (RMSE) on a hold-out test set containing 1,338 data points, \textit{i.e.}, 204,714 dipole values, is 0.14 (0.32) Debye. The accuracy of this model is comparable to our previous study, which could accurately represent the charge distribution in the methylenimmonium cation and ethylene.~\cite{Westermayr2020JCP}
The scatter plots that show the predicted dipole moment values against the reference values are shown in Fig.~\ref{fig:dipole}.

The dipole moments were fitted using 15,000 and 400 data points for training and validation, respectively. If not mentioned in the following, default model hyperparameters were used. A batch size of 64 was applied. The cutoff was set to 10 a.u and 1,024 features were used in combination with 5 interaction layers and 25 Gaussian functions to represent the molecule. The 3 hidden layers to map the tailored descriptors of the molecule to the dipole moment vectors contained 100 nodes each. 


%
%